\documentclass[a4paper]{spie}  %>>> use this instead for A4 paper
\usepackage[]{graphicx}

\title{Gaia on-board metrology: basic angle and best focus} 

\author{A. Mora\supit{a}\supit{b}, M. Biermann\supit{c}, A.G.A. Brown\supit{d}, D. Busonero\supit{e}, L. Carminati\supit{f}, J.M. Carrasco\supit{g}, F. Chassat\supit{f}, M. Erdmann\supit{h}, W.L.M. Gielesen\supit{i}, C. Jordi\supit{g}, D. Katz\supit{j}, R. Kohley\supit{a}, L. Lindegren\supit{k}, W. Loeffler\supit{c}, O. Marchal\supit{j}, P. Panuzzo\supit{j}, G. Seabroke\supit{l}, J. Sahlmann\supit{a}, E. Serpell\supit{m}\supit{n}, I. Serraller\supit{a}\supit{o}, F. van Leeuwen\supit{p}, W. van Reeven\supit{a}\supit{b}, T.C. van den Dool\supit{i} and L.L.A. Vosteen\supit{i}
\skiplinehalf
\supit{a}ESA-ESAC Gaia SOC, P.O. Box 78, 28691 Villanueva de la Ca\~{n}ada, Madrid, Spain; \\
\supit{b}Aurora Technology, Crown Business Centre, Heereweg 345, 2161 CA Lisse, The Netherlands; \\
\supit{c}Astronomisches Rechen-Institut, Moenchhofstr. 12-14, 69120 Heidelberg, Germany; \\
\supit{d}Leiden Observatory, Leiden University, P.O. Box 9513, 2300 RA, Leiden, The Netherlands; \\
\supit{e}Istituto Nazionale di Astrofisica - Osservatorio Astrofisico di Torino, V. Osservatorio 20, 10025 Pino T.se (TO), Italy; \\
\supit{f}Airbus Defence and Space, 31 rue des Cosmonautes, Z.I. du Palays, 31402 Toulouse Cedex 4, France \\
\supit{g}Departament d'Astronomia i Meteorologia, Institut del Ci\`encies del Cosmos (ICC), Universitat de Barcelona (IEEC-UB), c/ Martí i Franqu{\`e}s, 1, 08028, Barcelona, Spain;  \\
\supit{h}ESA-ESTEC Gaia Project Office, Keplerlaan 1, 2201 AZ Noordwijk, The Netherlands; \\
\supit{i}TNO Science and Industry, Stieltjesweg 1, 2600 AD Delft, The Netherlands; \\
\supit{j}GEPI, Observatoire de Paris, CNRS, Université Paris Diderot, 5 Place Jules Janssen, 92190, Meudon, France; \\
\supit{k}Lund Observatory, Department of Astronomy and Theoretical Physics, Lund University, Box 43, 22100, Lund, Sweden; \\
\supit{l}Mullard Space Science Laboratory, University College London, Holmbury St Mary, Dorking, Surrey RH5 6NT, UK; \\
\supit{m}ESA-ESOC Gaia Operations, Robert-Bosch-Strasse 5, 64293 Darmstadt, Germany; \\
\supit{n}Telespazio VEGA Deutschland GmbH, Europaplatz 5, 64293 Darmstadt, Germany; \\
\supit{o}GMV C/ Isaac Newton, 11, 28760 Tres Cantos, Madrid, Spain; \\
\supit{p}Institute of Astronomy, University of Cambridge, Madingley Road, Cambridge CB3OHA, UK; \\
}

\authorinfo{Further author information: (Send correspondence to A.M.)\\A.M.: E-mail: alcione.mora@esa.int, Telephone: +34 91 813 1480}
\begin{document} 
  \maketitle 

%%%%%%%%%%%%%%%%%%%%%%%%%%%%%%%%%%%%%%%%%%%%%%%%%%%%%%%%%%%%% 
\begin{abstract}
The Gaia payload ensures maximum passive stability using a single material, SiC, for most of its elements.
Dedicated metrology instruments are, however, required to carry out two functions: monitoring the basic angle
and refocusing the telescope. Two interferometers fed by the same laser are used to measure the basic angle changes at the level of $\mu$as (prad, micropixel), which is the highest level ever achieved in space. Two Shack-Hartmann wavefront
sensors, combined with an ad-hoc analysis of the scientific data are used to define and reach the overall best-focus.
In this contribution, the systems, data analysis, procedures and performance achieved during commissioning are
presented
\end{abstract}

%>>>> Include a list of keywords after the abstract 

\keywords{Astrometry, Gaia, metrology, interferometry, basic angle monitor, wavefront sensor, Shack-Hartmann, wavefront reconstruction, centroid, Cram\'er-Rao, spectral resolution}

%%%%%%%%%%%%%%%%%%%%%%%%%%%%%%%%%%%%%%%%%%%%%%%%%%%%%%%%%%%%%

\section{Introduction}

The ESA Gaia mission will provide astrometry of a billion objects in the Galaxy with unprecedent precision and accuracy. In addition, intermediate resolution spectra will be obtained for millions of sources. More details on the mission general goals can be found elsewhere\cite{LL:ESA-SCI(2000)4,2010SPIE.7731E..35D}. An overview of the commissioning results are also provided by Prusti [9143-503] (this conference).

The payload is composed of two off-axis telescopes sharing a common focal plane. Both telescopes have the same optical design: three mirror anastigmatic rectangular aperture. Four additional plane mirrors (three per telescope) are required to combine and fold the beams in the common focal plane. Two prisms are used to provide low resolution spectrophotometry, together with an afocal intermediate resolution Radial Velocity Spectrometer.

The whole system has been designed with extreme stability as a key feature. In this way, the whole payload is produced using a single material: silicon carbide (SiC), which provides exceptional rigidity and strength, low weight and high thermal conductivity. However, it is not enough to fulfil two stringent requirements, which define the two main topics of this contribution. First, Gaia must provide almost diffraction limited image performance in the visible, which prevents the telescopes to be perfectly aligned on-ground. Second, the basic angle (chief ray angular difference) between both telescopes must be tracked within 0.5 microarcsec (2.4 prad). Two metrology systems have been developed to fulfil those requirements: the Basic Angle Monitor (BAM, see Sect.~\ref{sect:bam}) and the two on-board Shack-Hartmann WaveFront Sensors (WFS). The latter are complemented by a dedicated analysis of the scientific data to define and obtain the best focus (see Sect.~\ref{sect:bestFocus}).

\section{The basic angle and the BAM}
\label{sect:bam}

\begin{figure}
  \begin{center}
	  \includegraphics[height=5.1cm]{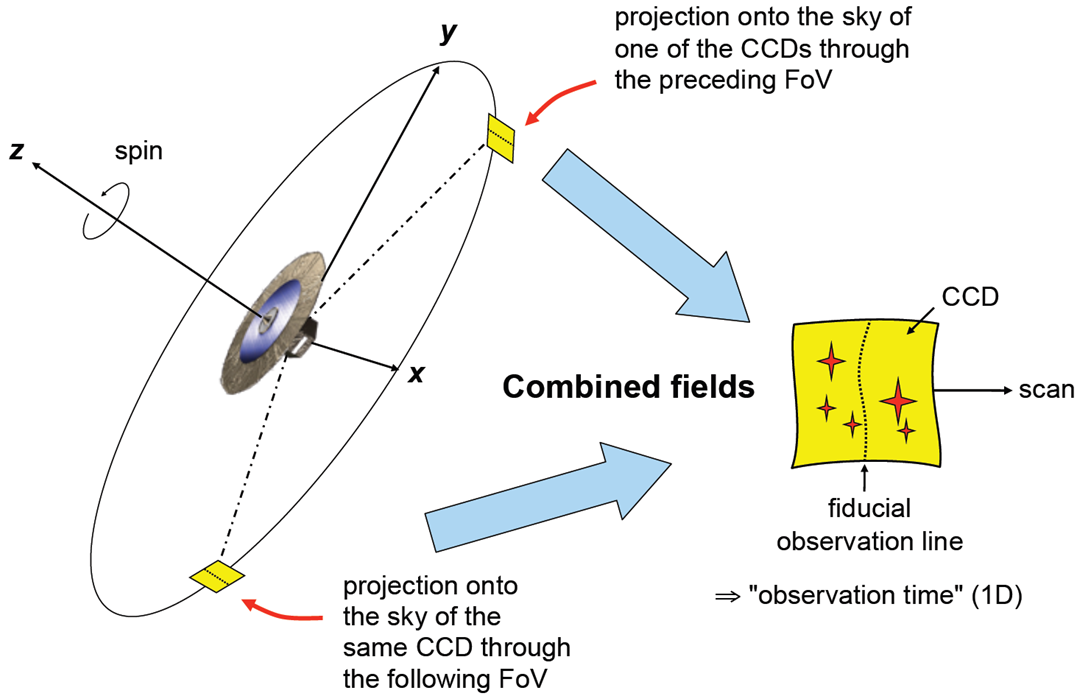}
	  \includegraphics[height=5.1cm]{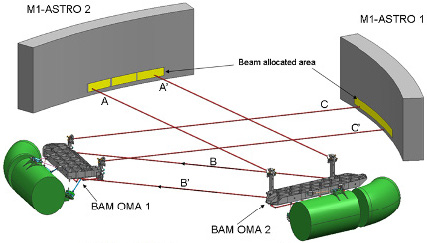}
  \end{center}
\caption{Left: Gaia working principle. Two telescopes separated by the large basic angle $\Gamma = 106.5^\circ$ produce stellar images in a combined focal plane. Image credit~\cite{2011EAS....45..109L}. Right: the BAM is a laser interferometer that injects two beams in each telescope entrance pupil. Image: Airbus Defence \& Space.
\label{fig:gaiaWorkingPrinciple}}
\end{figure}

Gaia aims at global astrometry (reference frame, stellar motions and parallaxes) at $\mu$as accuracy. The Gaia payload is composed of two telescopes scanning portions of the sky separated by the basic angle $\Gamma = 106.5^\circ$. A beam combiner is used to merge the images of both telescopes in the same focal plane (see Fig.~\ref{fig:gaiaWorkingPrinciple}) and produce the images in a combined focal plane. The fundamental principle is that differences in time acquisition between the stars of different focal planes can be translated into angular measurements.

All these measurements are affected if the basic angle is variable. Either it needs to be stable, or their variations known to the mission accuracy level ($\sim 1 \mu$as). Gaia is largely self-calibrating (calibration parameters estimated from observations). Therefore, low frequency variations ($f < 1 / 2P_{\rm rot}$) can be fully eliminated by self-calibration. High frequency random variations are also not a concern, because they are averaged during all transits.

However, high frequency systematic variations synchronized with spacecraft spin are a serious problem. They can only by partially eliminated by self-calibration and the residuals could create systematic errors in the astrometric results. Thus, high-frequency changes need to be monitored by metrology.

\subsection{BAM working principle}

The Basic Angle Monitoring device is in charge of the telescope line of sight change differential measurements. It basically generates one artificial fixed star per telescope, introducing two collimated laser beams into the primary mirrors (see Fig.~\ref{fig:gaiaWorkingPrinciple}). The BAM is composed of two optical benches: bar \#1 and bar \#2, in charge of producing the interference pattern for telescopes 1 and 2, respectively. The input light for all four beam is introduced by a polarisation maintaining single mode optical fibre in bar \#2. A number of beam splitters, and mirrors is used to generate all four beams (See~\cite{2012SPIE.8442E..1RG} for further details). The Gaia telescopes then generate the image, which is an interference pattern due to the coherent input light source. The relative Along Scan (AL) centroid displacements are then a direct measurement of the basic angle variations.

One reason why the artificial stars are interference patterns instead of point-like is because the required single image centroiding precision is much higher than for an unsaturated bright star. The sinusoidal BAM image can accommodate much more electrons than a point-like source, which translates into a much higher centroiding precision.

Several design rules have been implemented to ensure the BAM measures real changes in the line of sight, and not just its own instabilities. In particular:

\begin{itemize}
\item Insensitive to translation of bar \#1: the beams feeding bar \#1 are parallel
\item Insensitive to to rotation of bar \#1 along spin axis: same input/output beam separation.
\item Insensitive to different temperatures between bars: the optical path difference has been adjusted to make input/output planes to bar \#1 wavefronts.
\item Insensitive to laser beam point source motion: same light source for all beams
\item OPD $\sim$ 0, white light fringe must be in the pattern: the whole system OPD has been adjusted accordingly
\end{itemize}

\subsection{BAM data analysis
\label{sect:bamDataAnalysis}}

Several strategies have been proposed to analysed the BAM data. amongst them, cross-correlation, Fourier transform and direct fit. Each one has pros and cons.

Cross-correlation compares each observed BAM image to a reference template pattern. In this case, the cross-correlation function has not a single peak, as for stellar images, but is a periodic function. This is expected, because a dephase of a full period will still provide a good correlation. The central peak can be fitted with a sinusoidal function, the phase providing the required image shift with respect to the template. This algorithm is very fast and provides good precision. However, it provides little flexibility, and is thus easily affected by systematic errors.

Fourier transform is a natural choice for an interference pattern, which is at first order a sinusoidal function. It is still a fast option (although less than cross-correlation) and provides good precision, but is still affected by systematic errors and cannot be tailored to the particularities of the BAM images.

Finally, a mathematical model can be used to represent the BAM image, which is then fitted using a least squares algorithm. One free parameter will be the fringe shift, and the others are just nuisance variables. Maximum likelihood performance can be implemented using appropriate weights for each pixel. This method is much slower than the previous ones. On the other hand, it has, in principle, infinite flexibility to provide an accurate representation of the BAM pattern. In practice, an equilibrium is required to provide results in a reasonable time and avoid overfitting.

Direct fit modelling has been chosen for the Gaia regular BAM processing. It uses an analytic model inspired by Airbus Defence \& Space early studies, and has a physical meaning. It consists of the interference of two perfect Gaussian beams, represented by the centre coordinates, waist size and peak intensity. It is complemented by an additive constant, the fringe period and white light fringe, which is a straight line characterised by the position in the focal plane and the angle with respect to the vertical. This model provides reasonably fast computation but does not consider optical aberrations. The derivatives can also be explicitly computed. The total number of variables is then up to 12, although some can be kept fixed to speed up calculations or to provide a more physical fit. The noise model uses two components: Poisson shot noise and CCD read-out noise.

\begin{figure}
\begin{center}
\includegraphics[width=0.49\hsize]{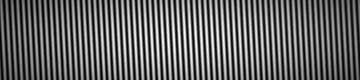}
\includegraphics[width=0.49\hsize]{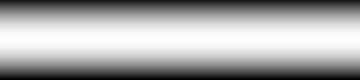}
\end{center}
\caption{BAM model images. Left: fringe interference pattern. Right TDI read-out background (there is no shutter). 
\label{fig:bamSynthetic}}
\end{figure}

The number of photons collected by a single pixel and its derivatives with respect to the 12 independent variables plus the white light fringe $y$-AC axis location (which is a fixed value, the middle point in the BAM pattern) are given in the following. SI units are used for all the variables in the equations (except for the sky brightness, which is given in electrons). The contribution to the interferogram and the background (TDI-readout + sky brightness) are shown separately. An example of model BAM data is shown in Fig.~\ref{fig:bamSynthetic}.

\subsubsection{Function values}

\begin{figure}
  \begin{center}
  \includegraphics[width=0.6\hsize]{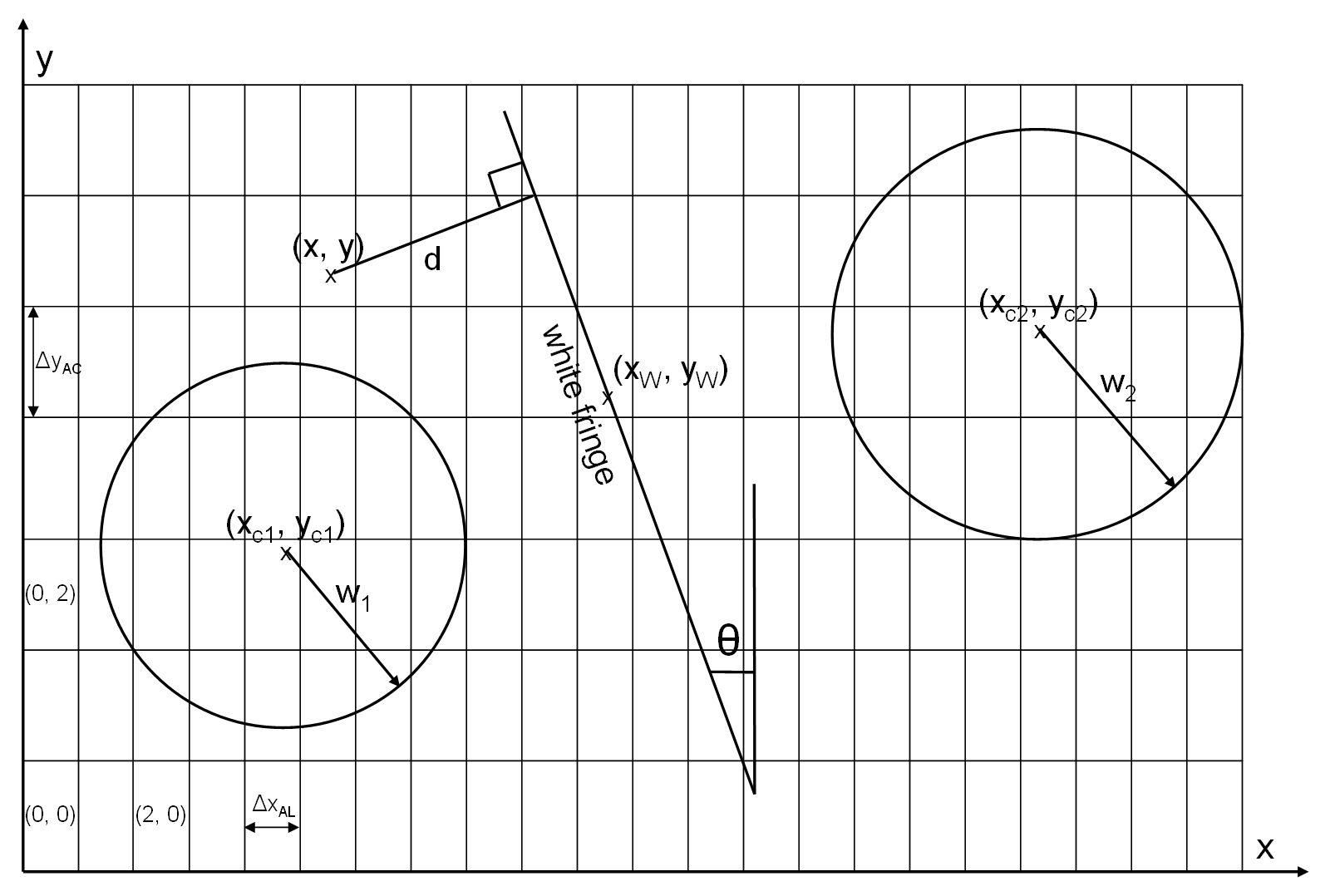}
  \end{center}
\caption{Generation of BAM images: pixelisation. The irradiance is integrated to obtain the number $N (i, j)$ of electrons collected per pixel $(i, j)$.
\label{fig:pixels}}
\end{figure}

The number of electrons $N$ and the background $B$ collected by each pixel $(i, j)$ during stare mode can be obtained integrating in $x$ and $y$. Fig.~\ref{fig:pixels} shows the model layout for pixel integration. 

\begin{equation}
  N(i,j) = \frac{ \Delta t_{\rm BAM} {\rm QE} }{ h \nu }
           \int_{ i \Delta x_{\rm AL} }^{ (i + 1) \Delta x_{\rm AL} } dx
           \int_{ j \Delta y_{\rm AC} }^{ (j + 1) \Delta y_{\rm AC} } dy
           \left({ I_{G1} + I_{G2} + 2 \sqrt{I_{G1} I_{G2}} \cos \delta }\right)
\label{eq:pixel_integration}
\end{equation}

\begin{equation}
  \begin{array}{ll}
    B (i,j)
            & \displaystyle = {\rm Sky} + \frac{ \Delta t_{\rm TDI} \Delta y_{\rm AC}  {\rm QE}}
                          { h \nu } \sqrt{ \frac{\pi}{2} } \\
            & \times \left[{ w_1 I_{G1} (x_{c1}, (j + 0.5) \Delta y_{\rm AC})
                          + w_2 I_{G2} (x_{c2}, (j + 0.5) \Delta y_{\rm AC}) }\right] \\
            & = {\rm Sky} + B_1 (i, j) + B_2 (i, j)
  \end{array}
\end{equation}

where

\begin{equation}
  I_G(x,y) = I_0 \exp \left({-2 \frac{ (x - x_c)^2 + (y - y_c)^2 }{w^2}}\right)
\end{equation}

\begin{equation}
  \delta (x, y) = 2 \pi \beta d = \frac{ 2 \pi B_I d }{ \lambda f }
\end{equation}

\begin{equation}
  d = \cos \theta (x - x_w) + \sin \theta (y - y_w)
\end{equation}

Note that with this formulation $|d|$ is the distance between a point $(x,y)$ and the white light fringe. Negative values are allowed for $d$ to make the derivatives simpler (the derivative of an absolute value is a piecewise-defined function). In fact, the sign of $d$ does not matter, because it is a multiplicative factor inside the argument of a cosine, which is an even function.

\subsubsection{Derivative: Gaussian peak intensity}

\begin{equation}
  \frac {\partial N(i,j)} {\partial I_{0, 1}} = 
           \frac{ \Delta t_{\rm BAM} {\rm QE} }{ h \nu }
           \int_{ i \Delta x_{\rm AL} }^{ (i + 1) \Delta x_{\rm AL} } dx
           \int_{ j \Delta y_{\rm AC} }^{ (j + 1) \Delta y_{\rm AC} } dy
           \left({ \frac{ I_{G1} + \sqrt{I_{G1} I_{G2}} \cos \delta }{I_{0, 1}} }\right)
\end{equation}

\begin{equation}
  \frac{\partial B (i,j)}{\partial I_{0,1}} = \frac { B_1 (i, j) } { I_{0,1} }
\end{equation}

\begin{equation}
  \frac {\partial N(i,j)} {\partial I_{0, 2}} = 
           \frac{ \Delta t_{\rm BAM} {\rm QE} }{ h \nu }
           \int_{ i \Delta x_{\rm AL} }^{ (i + 1) \Delta x_{\rm AL} } dx
           \int_{ j \Delta y_{\rm AC} }^{ (j + 1) \Delta y_{\rm AC} } dy
           \left({ \frac{ I_{G2} + \sqrt{I_{G1} I_{G2}} \cos \delta }{I_{0, 2}} }\right)
\end{equation}

\begin{equation}
  \frac{\partial B (i,j)}{\partial I_{0,2}} = \frac { B_2 (i, j) } { I_{0,2} }
\end{equation}

\subsubsection{Derivative: Gaussian beam waist}

\begin{equation}
  \begin{array}{ll}
    \displaystyle \frac {\partial N(i,j)} {\partial w_1} = 
      & \displaystyle \frac{ \Delta t_{\rm BAM} {\rm QE} }{ h \nu }
        \int_{ i \Delta x_{\rm AL} }^{ (i + 1) \Delta x_{\rm AL} } dx
        \int_{ j \Delta y_{\rm AC} }^{ (j + 1) \Delta y_{\rm AC} } dy \\
      & \displaystyle \left({ I_{G1} + \sqrt{I_{G1} I_{G2}} \cos \delta }\right)
        \left({ 4 \frac { (x - x_{c1})^2 + (y - y_{c1})^2 }{ w_1^3 } }\right)
  \end{array}
\end{equation}

\begin{equation}
  \frac{\partial B (i,j)}{\partial w_1} =
     B_1 (i, j) \left({
        \frac{1}{w_1} +
        \frac { 4 \left[{ (j + 0.5) \Delta y_{\rm AC} - y_{c1} }\right]^2 } { w_1^3 }
                      }\right)
\end{equation}

\begin{equation}
  \begin{array}{ll}
    \displaystyle \frac {\partial N(i,j)} {\partial w_2} = 
      & \displaystyle \frac{ \Delta t_{\rm BAM} {\rm QE} }{ h \nu }
        \int_{ i \Delta x_{\rm AL} }^{ (i + 1) \Delta x_{\rm AL} } dx
        \int_{ j \Delta y_{\rm AC} }^{ (j + 1) \Delta y_{\rm AC} } dy \\
      & \displaystyle \left({ I_{G2} + \sqrt{I_{G1} I_{G2}} \cos \delta }\right)
        \left({ 4 \frac { (x - x_{c2})^2 + (y - y_{c2})^2 }{ w_2^3 } }\right)
  \end{array}
\end{equation}

\begin{equation}
  \frac{\partial B (i,j)}{\partial w_2} =
     B_2 (i, j) \left({
        \frac{1}{w_2} +
        \frac { 4 \left[{ (j + 0.5) \Delta y_{\rm AC} - y_{c2} }\right]^2 } { w_2^3 }
                      }\right)
\end{equation}

\subsubsection{Derivatives: $xy$ location of Gaussian peaks}

\begin{equation}
  \begin{array}{ll}
    \displaystyle \frac {\partial N(i,j)} {\partial x_{c1}} = 
      & \displaystyle \frac{ \Delta t_{\rm BAM} {\rm QE} }{ h \nu }
        \int_{ i \Delta x_{\rm AL} }^{ (i + 1) \Delta x_{\rm AL} } dx
        \int_{ j \Delta y_{\rm AC} }^{ (j + 1) \Delta y_{\rm AC} } dy \\
      & \displaystyle \left({ I_{G1} + \sqrt{I_{G1} I_{G2}} \cos \delta }\right)
        \left({ \frac { 4 (x - x_{c1}) }{ w_1^2 } }\right)
  \end{array}
\end{equation}

\begin{equation}
  \frac{\partial B (i,j)}{\partial x_{c1}} = 0
\end{equation}

\begin{equation}
  \begin{array}{ll}
    \displaystyle \frac {\partial N(i,j)} {\partial x_{c2}} = 
      & \displaystyle \frac{ \Delta t_{\rm BAM} {\rm QE} }{ h \nu }
        \int_{ i \Delta x_{\rm AL} }^{ (i + 1) \Delta x_{\rm AL} } dx
        \int_{ j \Delta y_{\rm AC} }^{ (j + 1) \Delta y_{\rm AC} } dy \\
      & \displaystyle \left({ I_{G2} + \sqrt{I_{G1} I_{G2}} \cos \delta }\right)
        \left({ \frac { 4 (x - x_{c2}) }{ w_2^2 } }\right)
  \end{array}
\end{equation}

\begin{equation}
  \frac{\partial B (i,j)}{\partial x_{c2}} = 0
\end{equation}

\begin{equation}
  \begin{array}{ll}
    \displaystyle \frac {\partial N(i,j)} {\partial y_{c1}} = 
      & \displaystyle \frac{ \Delta t_{\rm BAM} {\rm QE} }{ h \nu }
        \int_{ i \Delta x_{\rm AL} }^{ (i + 1) \Delta x_{\rm AL} } dx
        \int_{ j \Delta y_{\rm AC} }^{ (j + 1) \Delta y_{\rm AC} } dy \\
      & \displaystyle \left({ I_{G1} + \sqrt{I_{G1} I_{G2}} \cos \delta }\right)
        \left({ \frac { 4 (y - y_{c1}) }{ w_1^2 } }\right)
  \end{array}
\end{equation}

\begin{equation}
  \frac{\partial B (i,j)}{\partial y_{c1}} =
    B_1 (i, j) \frac{4}{w_1^2} [(j + 0.5) \Delta y_{\rm AC} - y_{c1}]
\end{equation}

\begin{equation}
  \begin{array}{ll}
    \displaystyle \frac {\partial N(i,j)} {\partial y_{c2}} = 
      & \displaystyle \frac{ \Delta t_{\rm BAM} {\rm QE} }{ h \nu }
        \int_{ i \Delta x_{\rm AL} }^{ (i + 1) \Delta x_{\rm AL} } dx
        \int_{ j \Delta y_{\rm AC} }^{ (j + 1) \Delta y_{\rm AC} } dy \\
      & \displaystyle \left({ I_{G2} + \sqrt{I_{G1} I_{G2}} \cos \delta }\right)
        \left({ \frac { 4 (y - y_{c2}) }{ w_2^2 } }\right)
  \end{array}
\end{equation}

\begin{equation}
  \frac{\partial B (i,j)}{\partial y_{c2}} =
    B_2 (i, j) \frac{4}{w_2^2} [(j + 0.5) \Delta y_{\rm AC} - y_{c2}]
\end{equation}

\subsubsection{Derivative: $xy$ location of white light fringe}

\begin{equation}
  \frac {\partial N(i,j)} {\partial x_w } = 
           \frac{ \Delta t_{\rm BAM} {\rm QE} }{ h \nu }
           \int_{ i \Delta x_{\rm AL} }^{ (i + 1) \Delta x_{\rm AL} } dx
           \int_{ j \Delta y_{\rm AC} }^{ (j + 1) \Delta y_{\rm AC} } dy
           \left({
             2 \sqrt{I_{G1} I_{G2}} \sin \delta \frac { 2 \pi B_I }{ \lambda f } \cos \theta
                 }\right)
\end{equation}

\begin{equation}
  \frac{\partial B (i,j)}{\partial x_w} = 0
\end{equation}

The white light fringe $y$-AC axis location is not considered an independent variable in this technical note. The following equations are only given for the sake of completeness.

\begin{equation}
  \frac {\partial N(i,j)} {\partial y_w } = 
           \frac{ \Delta t_{\rm BAM} {\rm QE} }{ h \nu }
           \int_{ i \Delta x_{\rm AL} }^{ (i + 1) \Delta x_{\rm AL} } dx
           \int_{ j \Delta y_{\rm AC} }^{ (j + 1) \Delta y_{\rm AC} } dy
           \left({
             2 \sqrt{I_{G1} I_{G2}} \sin \delta \frac { 2 \pi B_I }{ \lambda f } \sin \theta
                 }\right)
\end{equation}

\begin{equation}
  \frac{\partial B (i,j)}{\partial y_w} = 0
\end{equation}

\subsubsection{Derivative: white light fringe angle with respect to $y$-AC axis}

\begin{equation}
  \begin{array}{ll}
    \displaystyle \frac {\partial N(i,j)} {\partial \theta } = 
      & \displaystyle \frac{ \Delta t_{\rm BAM} {\rm QE} }{ h \nu }
        \int_{ i \Delta x_{\rm AL} }^{ (i + 1) \Delta x_{\rm AL} } dx
        \int_{ j \Delta y_{\rm AC} }^{ (j + 1) \Delta y_{\rm AC} } dy \\
      & \displaystyle 2 \sqrt{I_{G1} I_{G2}} \sin \delta
        \frac { 2 \pi B_I }{ \lambda f }
        [ \sin \theta (x - x_w) - \cos \theta (y - y_w) ]
  \end{array}
\end{equation}

\begin{equation}
  \frac{\partial B (i,j)}{\partial \theta} = 0
\end{equation}

\subsubsection{Derivative: wavelength}

\begin{equation}
  \frac {\partial N(i,j)} {\partial \lambda} = 
           \frac{ \Delta t_{\rm BAM} {\rm QE} }{ h \nu }
           \int_{ i \Delta x_{\rm AL} }^{ (i + 1) \Delta x_{\rm AL} } dx
           \int_{ j \Delta y_{\rm AC} }^{ (j + 1) \Delta y_{\rm AC} } dy
           \left({
             2 \sqrt{I_{G1} I_{G2}} \sin \delta \frac { 2 \pi B_I d }{ \lambda^2 f }
                 }\right)
\end{equation}

\begin{equation}
  \frac{\partial B (i,j)}{\partial \lambda} = 0
\end{equation}

\subsubsection{Derivative: Sky brightness}

\begin{equation}
  \frac {\partial N(i,j)} {\partial {\rm Sky}} = 0
\end{equation}

\begin{equation}
  \frac{\partial B (i,j)}{\partial {\rm Sky}} = 1
\end{equation}

\subsubsection{Conversion factors: non-SI units}

non-SI units are more convenient to understand the value of some variables: $xy$ location on the focal plane (pixels), Gaussian peak irradiance (electrons collected per pixel without considering interference) and white light fringe angle (degrees). The derivatives with respect to these non-SI units can be computed following the chain rule for derivatives:

\begin{equation}
  \frac {\partial N} {\partial x_{\rm pix}} 
  = \frac {\partial N} {\partial x} \Delta x_{\rm AL}
\end{equation}

\begin{equation}
  \frac {\partial N} {\partial y_{\rm pix}} 
  = \frac {\partial N} {\partial y} \Delta y_{\rm AC}
\end{equation}

\begin{equation}
  \frac {\partial N} {\partial {N_{G, \rm peak, e^-}}} 
  = \frac {\partial N} {\partial I_0} 
    \frac{ h \nu }
         { \Delta t_{\rm BAM} {\rm QE} \Delta x_{\rm AL} \Delta y_{\rm AC}  }
\end{equation}

\begin{equation}
  \frac {\partial N} {\partial \theta_{\rm deg}} 
  = \frac {\partial N} {\partial \theta} \frac{\pi}{180} 
\end{equation}

\begin{equation}
  \frac {\partial N} {\partial P_{\rm pix}} 
  = \frac {\partial N} {\partial \lambda} \frac{B \Delta x_{\rm AL}}{f}
\end{equation}

\subsubsection{MIT-IDT pipeline}

The BAM data is processed by two main systems after being downlinked: MIT and IDT (see also Riva et al. [9150-73], this conference, for additional BAM DPAC data processing).

MIT, the MOC Interface Task (see Siddiqui et al. [9149-91], this conference), reconstructs the telemetry stream, identifies the BAM spacecraft SP4 data packets and stores them in a data base. This process is sequential in nature, because of the different checks required to ensure the correct data assembly and integrity.

IDT, the Initial Data Treatment assembles the different SP4 telemetry packets and processes them into a high level object: the BamElementary, which is subsequently stored in a data base. The most CPU intensive operation is the latest one, which is thus highly parallelised using a fully automated Java pipeline. This system is always active, and is capable to process one day of data in a few hours. The only manual operations are software and calibration (BamStatus) updates. The ESA-ESAC DPCE cluster resources devoted to IDT are typically 8 nodes composed of 2 Intel X5550 CPUs, 8 cores in total running at 2.66 GHz sharing 32 GB of RAM.

\subsection{BAM real behaviour}

\begin{figure}
\begin{center}
\includegraphics[width=\hsize]{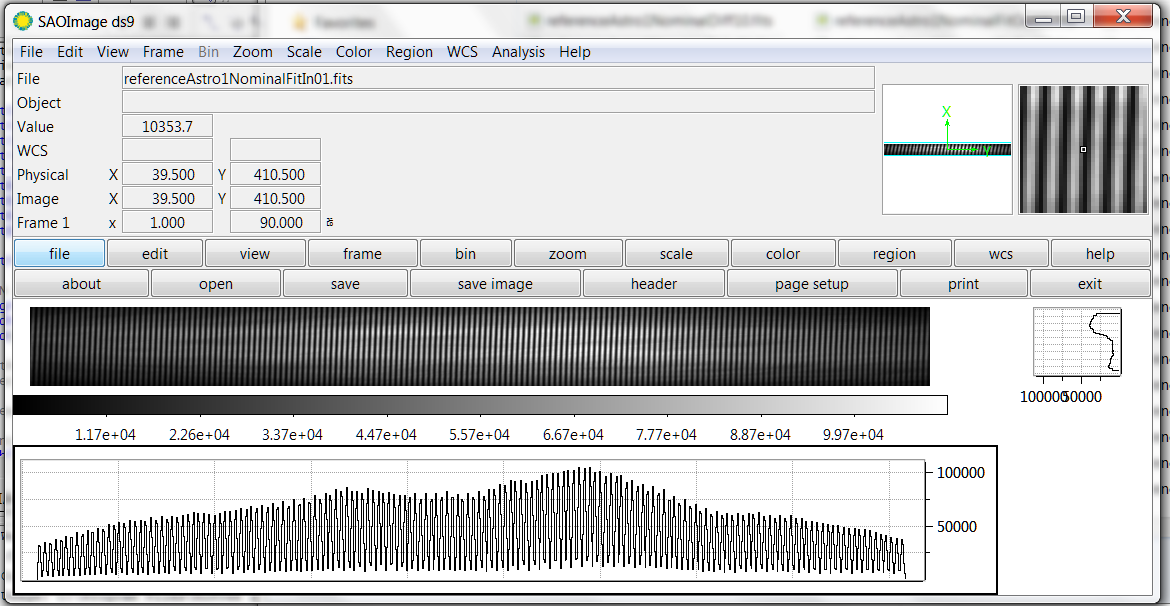}
\end{center}
\caption{Horizontal and vertical cross-sections for a telescope 1 BAM pattern. The non-Gaussianity of the fringe envelope is evident. Plenty of structure can be appreciated in the form of an additional low frequency interference pattern. Its origin is still uncertain.
\label{fig:bamNonGaussianity}}
\end{figure}

The real BAM images roughly resemble the idealised model proposed in Sect.~\ref{sect:bamDataAnalysis}. The most obvious difference is the deviation with respect to the idealised Gaussian envelope, Fig.~\ref{fig:bamNonGaussianity} shows a horizontal and vertical BAM profile for telescope 1. It is apparent that both cross-sections are not Gaussian. The ratio between the real and model patterns was interpolated, smoothed and stretched to real focal plane coordinates (the samples are rectangular, with a horizontal:vertical aspect ratio of 1:12). The resultant image is a sort of flat-field, which reveals plenty of structure resembling and additional low-frequency interference pattern overimposed on the main BAM fringe pattern. The origin of this additional interference is still unclear.

\begin{figure}
\begin{center}
\includegraphics[width=0.49\hsize]{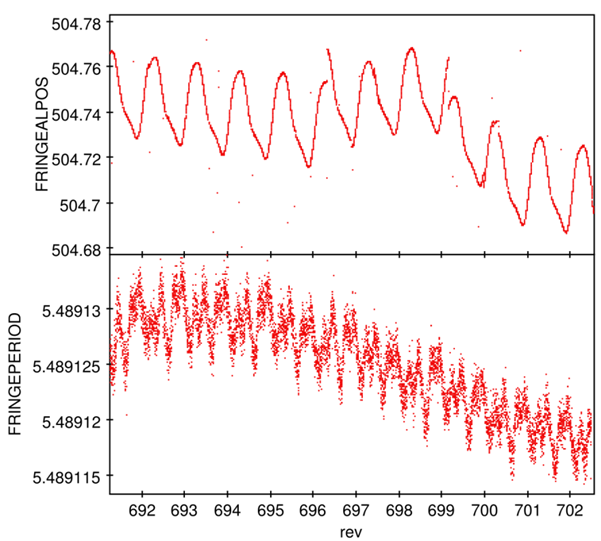}
\includegraphics[width=0.49\hsize]{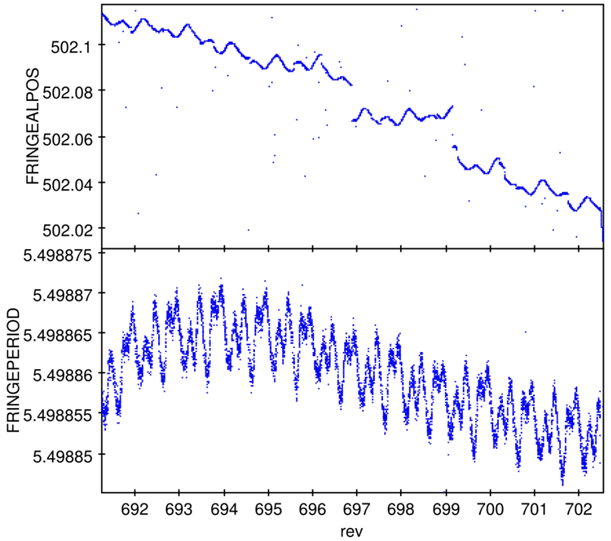}
\end{center}
\caption{BAM fringe location (top) and period (bottom) for telescope 1 (left) and 2 (right) covering a time interval of three days.
\label{fig:fringePhasePeriod}}
\end{figure}

The most important parameters provided by BAM-IDT for each pattern are the fringe location (phase) and period. Fig.~\ref{fig:fringePhasePeriod} shows the results for a particular three days interval. Four features can be identified: periodic, Sun synchronous, changes in the fringe phase, with amplitude $\sim$1mas, fringe phase discontinuities (several per day), fringe phase mid-long term evolution and fringe period variability.

\begin{figure}
\begin{center}
\includegraphics[height=5.5cm]{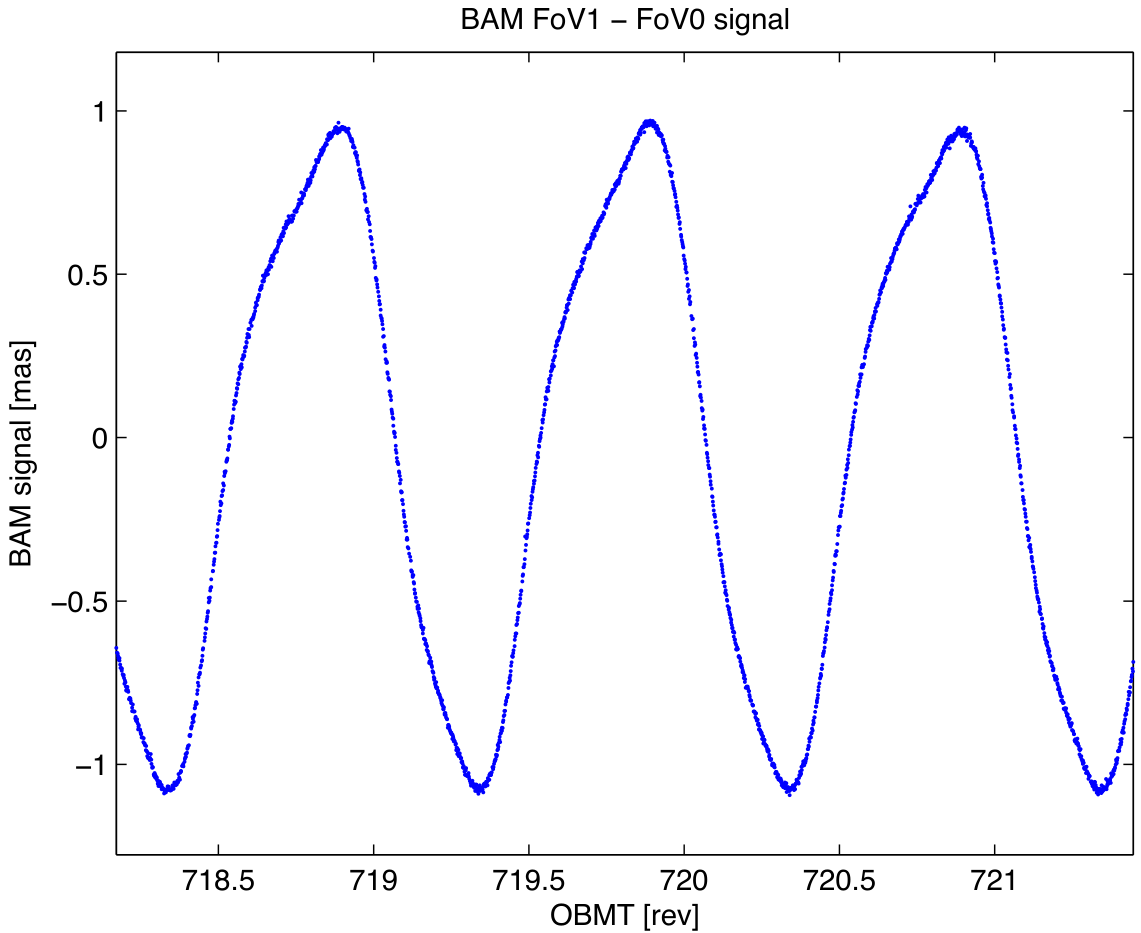}
\includegraphics[height=5.5cm]{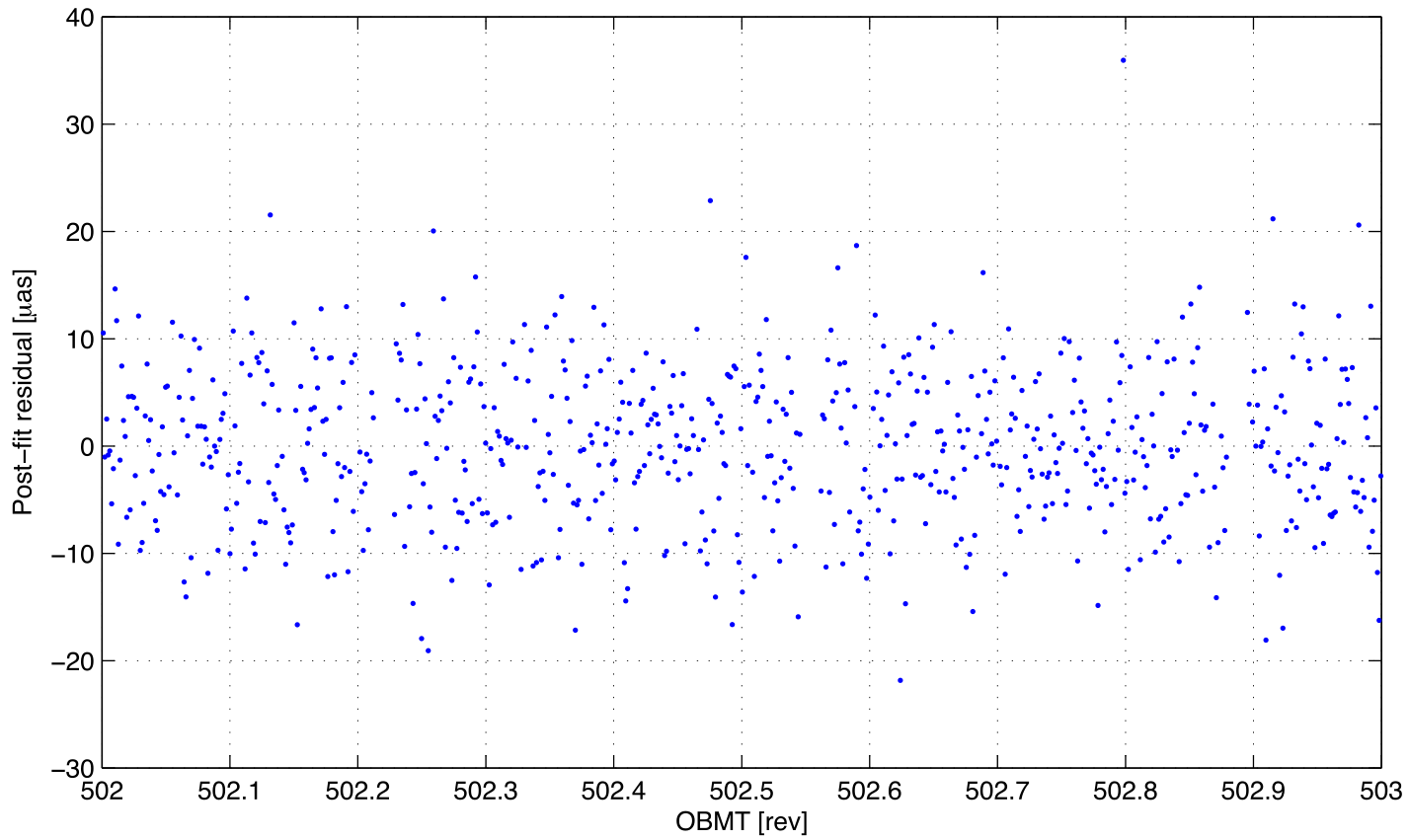}
\end{center}
\caption{Fourier analysis of the BAM fringe phase periodic signal. An expansion up to order 12 in the six hours rotation period provides fit residuals at the $\mu$as level, with no clear remaining systematic trends.
\label{fig:bamFourierAnalysis}}
\end{figure}

The periodic features could be very dangerous in terms of systematic errors, as discussed in Sect.~\ref{sect:bam}, if they represent a physical change of the line of sight and are not accounted for by the astrometric solution. A Fourier analysis has thus been carried out on the periodic signal. It has shown that the periodicity can be well approximated by a Fourier expansion up to order 12 in the six hours rotation period. The fit residuals are very small, at the $\mu$as level, with no clear systematic trends apparent (see Fig.~\ref{fig:bamFourierAnalysis}). The temporal evolution of the Fourier coefficients is very smooth. This means that the effect will be accurately characterised, and could be efficiently handled by the astrometric solution.

\begin{figure}
\begin{center}
\includegraphics[width=0.95\hsize]{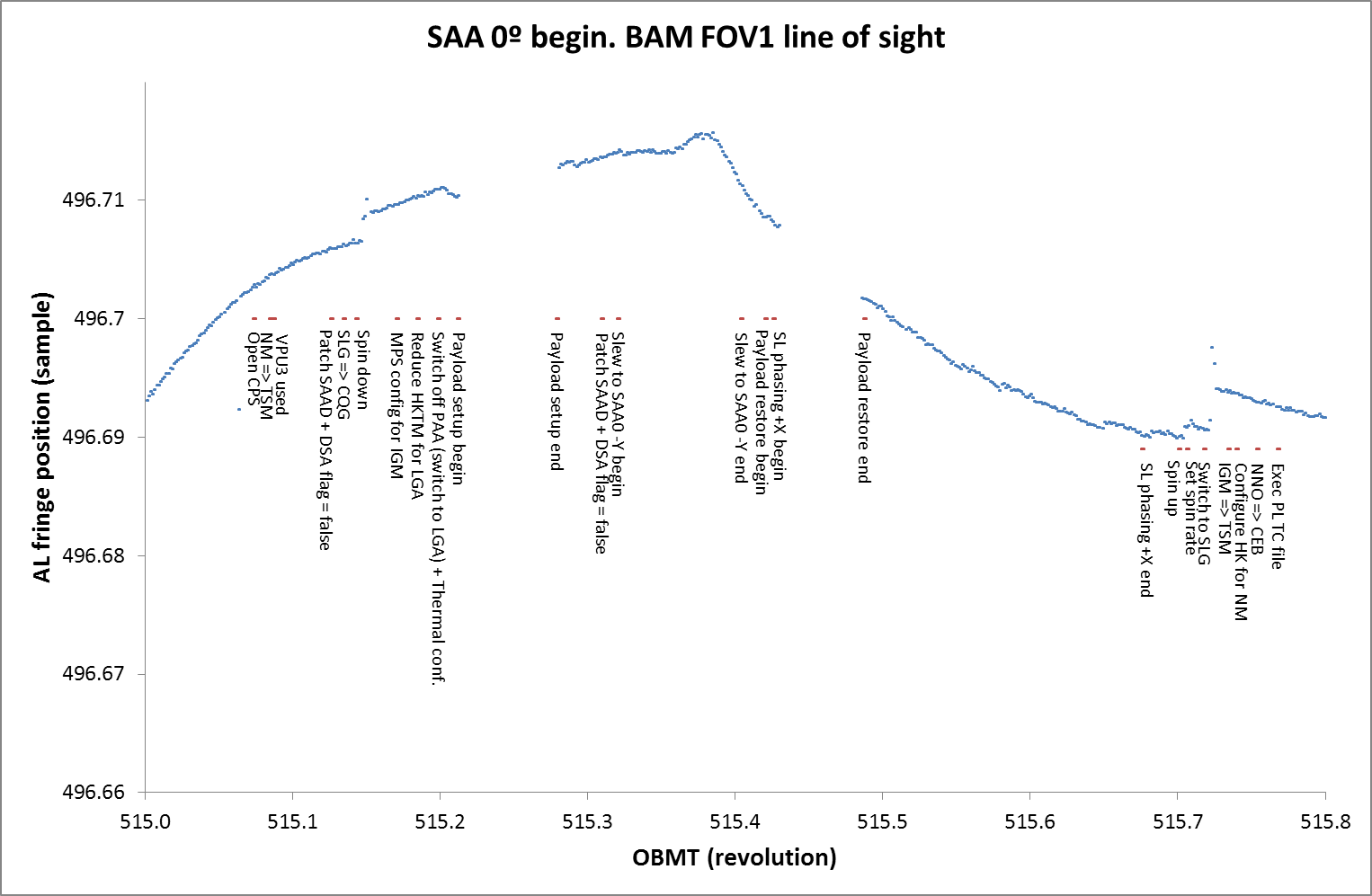}

\vspace{0.5cm}
\includegraphics[width=0.95\hsize]{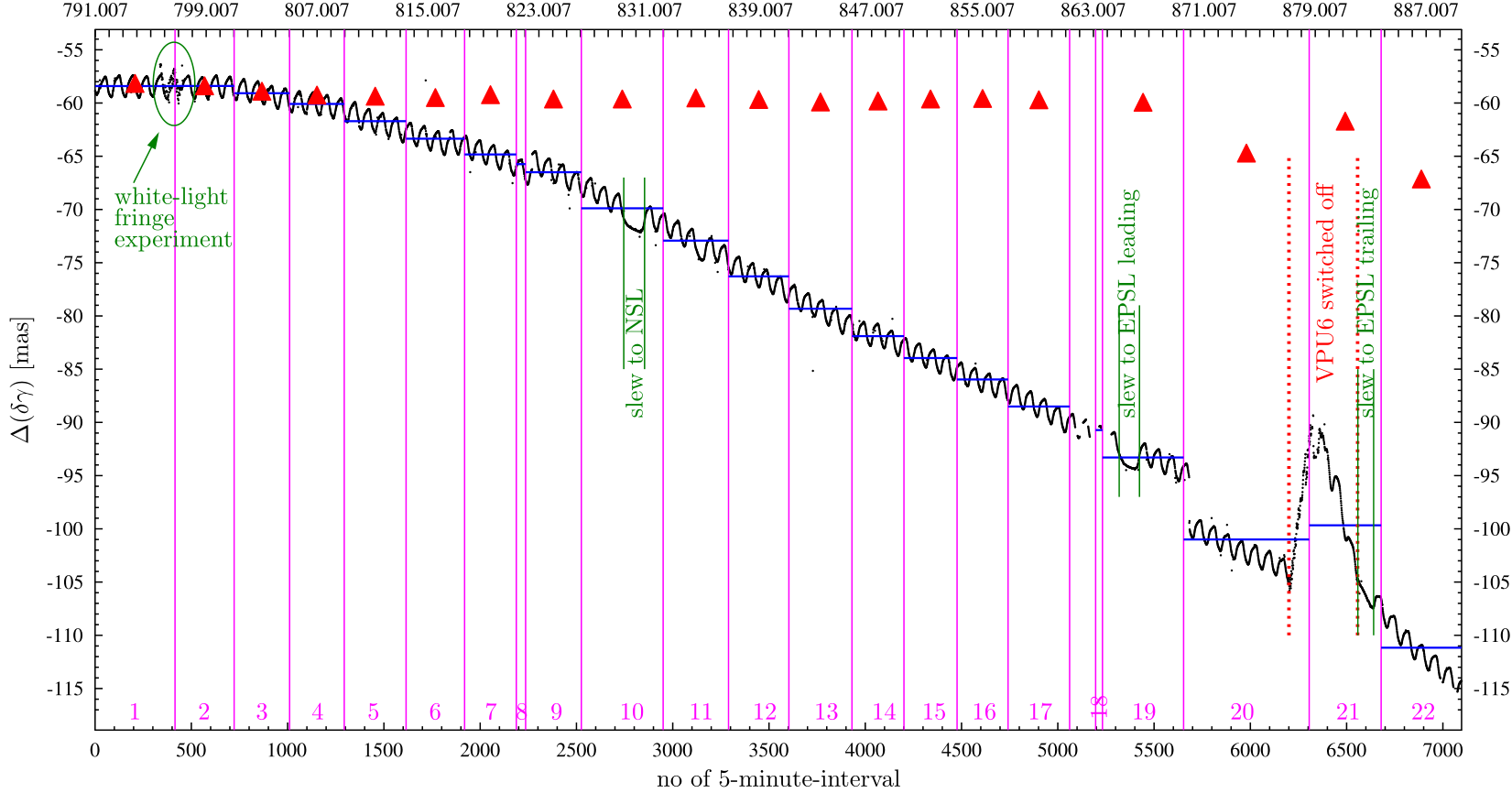}
\end{center}
\caption{BAM fringe phase discontinuities. Top, several line of sight changes per day are identified by the BAM. The largest ones can typically be traced to disrupting on-board activities. Bottom, the One Day Astrometric Solution (ODAS, red triangles) was able to verify the changes in basic angle measured by the BAM (black line) were real using stellar measurements. 
\label{fig:bamDiscontinuities}}
\end{figure}

Many discontinuities have been identified in the BAM fringe location signal. The amplitude goes from the pixel level ($\sim$50 mas) to sub-mas. The large amplitude ones are typically related to on-board disrupting activities such as spin-up and down or station keeping manoeuvres (see Fig.~\ref{fig:bamDiscontinuities}). The Gaia astrometric solution has provision to handle discontinuities. However, the proof that the discontinuities in the BAM signal came from telescope line of sight changes came from the cross-comparison with the One Day Astrometric Solution (ODAS). This analysis is carried out daily in a thin ring in the sky to diagnose problems on-board, is based in the observation of stars, and is insensitive to the periodic oscillations. However, it was able to match large jumps in the BAM signal to jumps in the stellar basic angle, confirming the BAM is measuring a real signal. The slow day-to-day fringe phase evolution was identified as false. Long term BAM stability is not required, though, as discussed in Sect.~\ref{sect:bam}.

\begin{figure}
\begin{center}
\includegraphics[width=0.65\hsize]{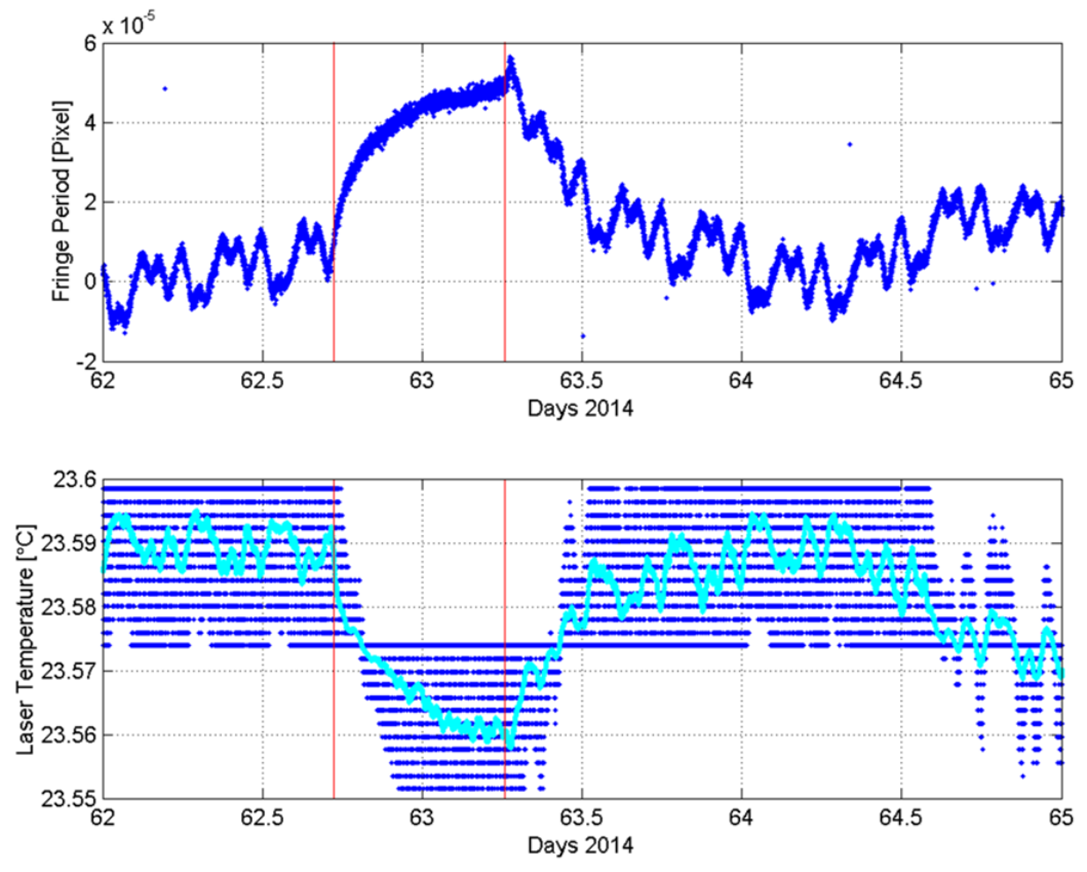}
\end{center}
\caption{BAM fringe period instability. Pseudo-periodic changes in the fringe period are correlated to small amplitude changes in the laser temperature, which in turn are related to the mode change of CCDs when the telescopes cross high stellar density regions.
\label{fig:bamFringePeriodInstability}}
\end{figure}

Surprisingly enough, the fringe period variability has been traced to quasi-periodic changes of the laser temperature with an amplitude of $\sim$0.005 K (see Fig.~\ref{fig:bamFringePeriodInstability}). These small changes introduce periodic shifts in the fringe period at the level of $\sim$1/250,000. The origin of these temperature changes is related to the operation of the some CCDs in the focal plane, collecting spectra, whose mode switches between low and high resolution more often when the stellar density increases (i.e., for galactic plane crossings). This effect is undesirable, but predictable. Mitigation schemes are also under study.

\begin{figure}
\begin{center}
\includegraphics[width=0.49\hsize]{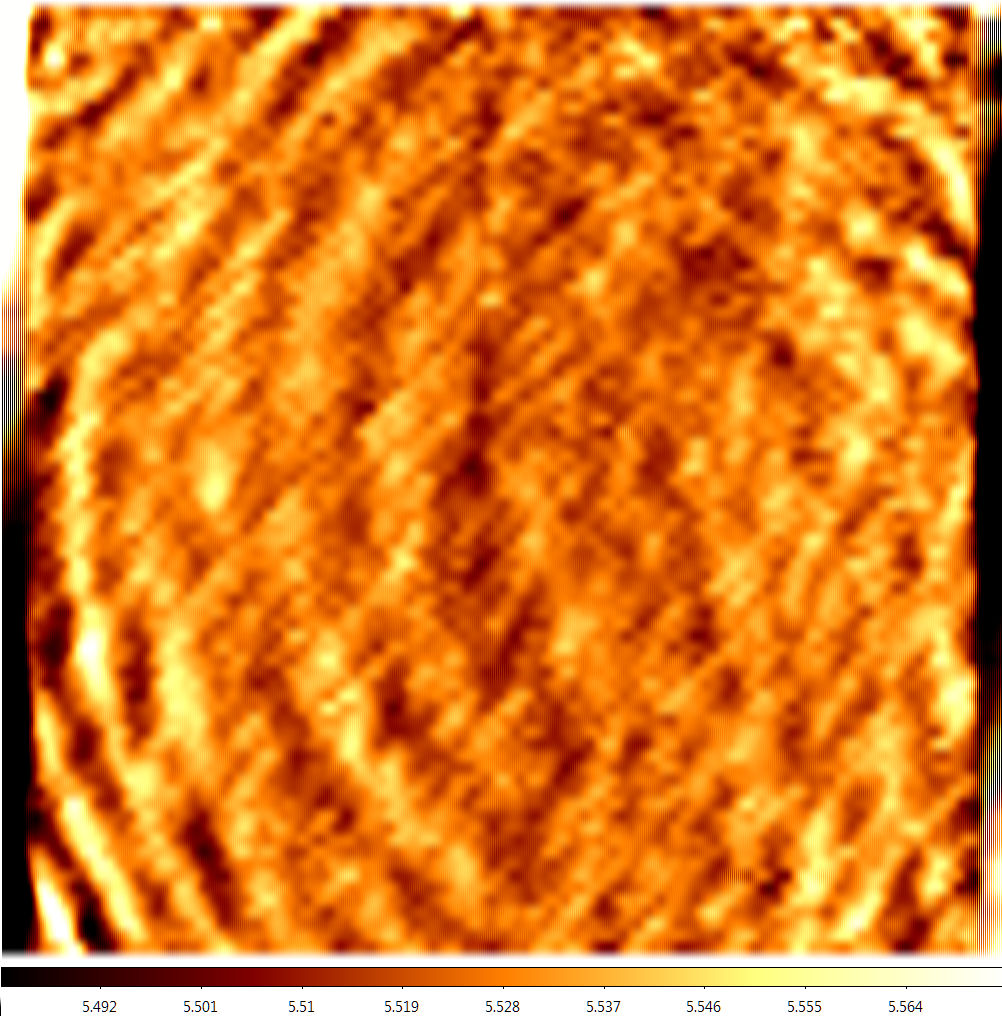}
\includegraphics[width=0.49\hsize]{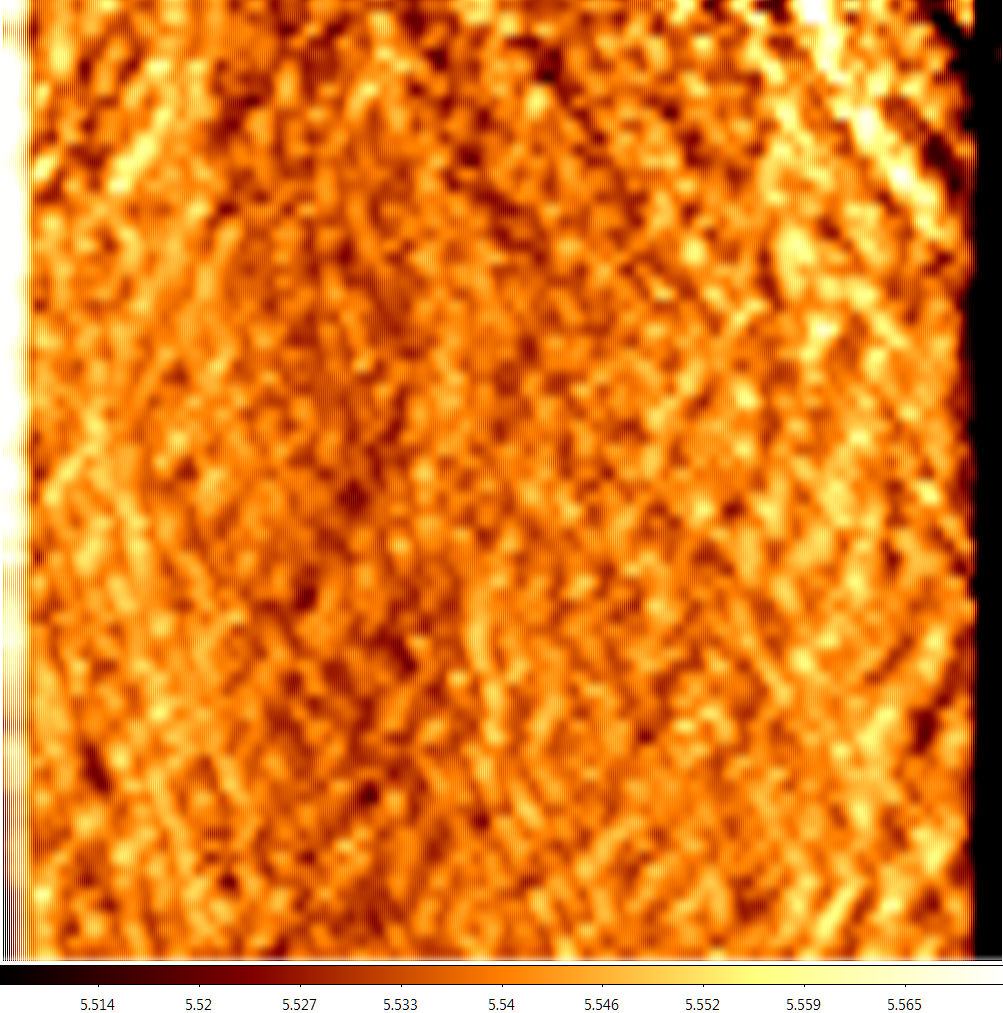}
\end{center}
\caption{
The wavelet transform of BAM nominal was used to obtain the spatial variation of the fringe period. The results for telescope 1 (left) and 2 (right) show that the fringe period is not constant, and depends on the position within the pattern. The fringes are thus not purely plane-parallel lines.
\label{fig:bamWaveletTransform}}
\end{figure}

Finally, a Morlet continuous wavelet transform analysis has been carried out on the BAM data. The immediate goal was to determine the white light fringe location. In addition, it revealed that the fringe period is not constant throughout the image (see Fig.~\ref{fig:bamWaveletTransform}). This means that the fringes cannot be accurately represented by evenly spaced plane-parallel lines. A better analysis that takes this effect into account is left for future work after commissioning.

\section{In-orbit realignment: best focus}
\label{sect:bestFocus}

\begin{figure}
\begin{center}
\includegraphics[height=4.3cm]{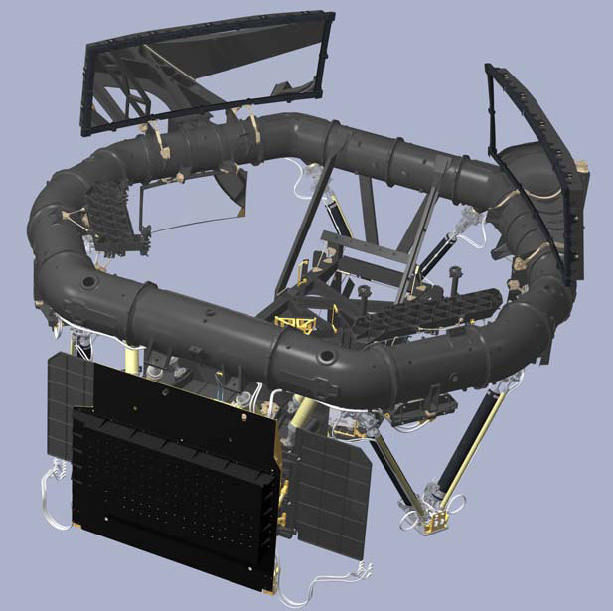}
\includegraphics[height=4.3cm]{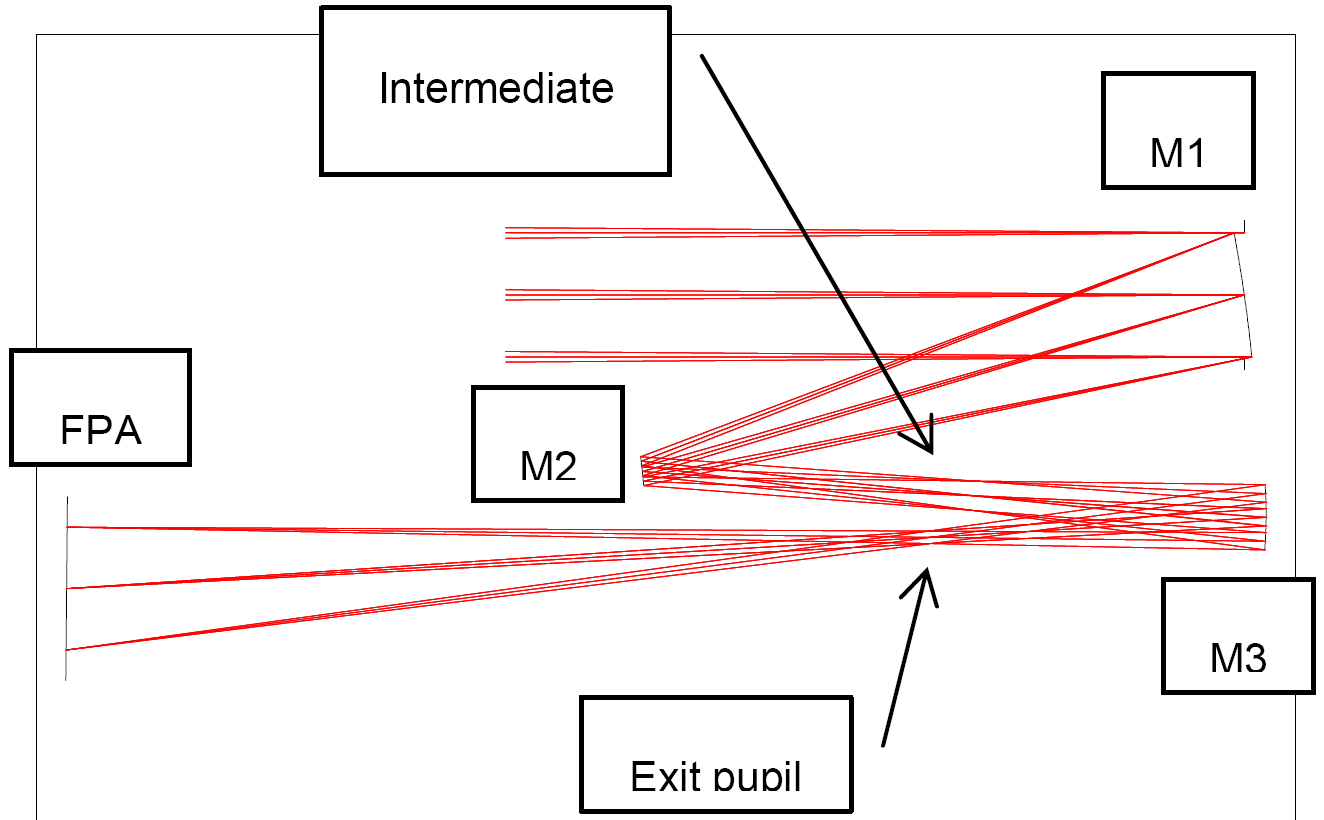}
\includegraphics[height=4.3cm]{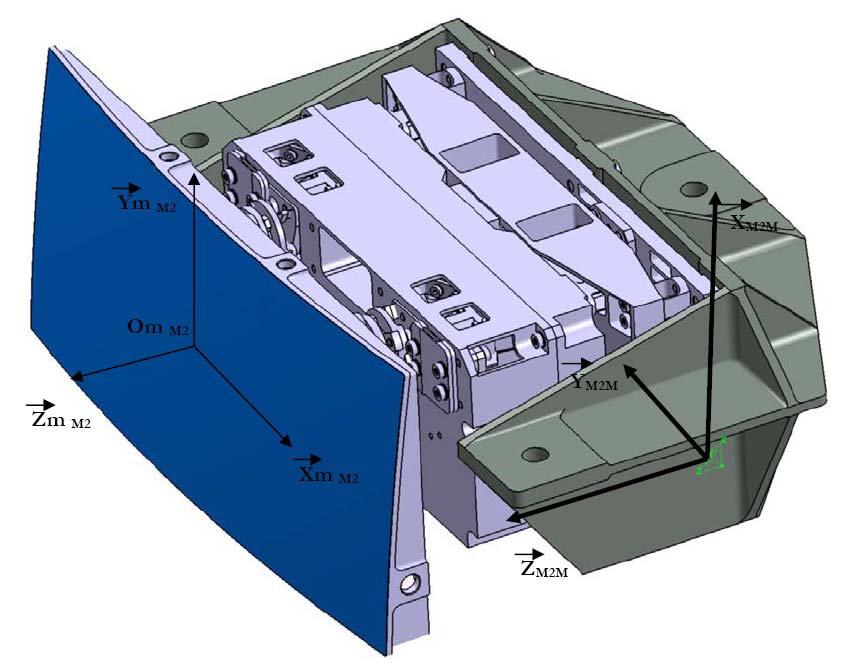}
\end{center}
\caption{Gaia payload overview (left), unfolded telescope optical design (centre) and M2 Movement Mechanism (M2MM, right). Courtesy Airbus Defence \& Space.}
\label{fig:gaiaOpticsWfs}
\end{figure}

The payload is composed of two twin off-axis three mirror anastigmatic telescopes (TMA) with a rectangular pupil of 1.45$\times$0.5 m feeding a common focal plane\cite{2010SPIE.7731E..35D} (see also~\cite{2012SPIE.8442E..1PK}). In addition to the powered surfaces, several plane mirrors are required, two for the pupil plane beam combiner and two for a common periscope. Fig.~\ref{fig:gaiaOpticsWfs} provides an overview of the overall optical system. Most of the payload, including the mirrors, focal plane and torus support structure are made of silicon carbide (SiC). This material combines low weight with high stiffness and thermal conductivity, providing a very homogeneous temperature in-orbit.

Gaia operates in the visible range (300-1050 nm) with a very high quality optical system (total wavefront error budget $\sim$50 nm). The mechanical tolerances for such a folded TMA system are tight, and smaller that the typical perturbations estimated for the launch vibrations and gravity release. Focusing mechanisms have thus been incorporated to move each secondary mirror (M2), the so called M2 Movement Mechanisms (M2MM). Each M2MM has a fully redundant set of actuators capable of orienting the M2 surface with five degrees of freedom (which is enough for a rotationally symmetric surface). Fig.~\ref{fig:gaiaOpticsWfs} also shows a model of the system.

The in-orbit telescope focusing has been a two step process. For the first iterations, two WaveFront Sensors (WFS) were used to correct most of the launch induced and gravity-release aberrations. A number of additional iterations were also carried out, where the WFS input was combined to the analysis of the scientific data. The final focus setting was a compromise position based on scientific criteria. Some aspects of the WFS performance are discussed in Sec.\ref{sect:wfs}, while an overview of the iterative best focus process and the metrics based on the scientific data are presented in Sec.\ref{sect:bestFocusMetrics}.

\subsection{The Gaia wavefront sensors
\label{sect:wfs}}

\begin{figure}
\begin{center}
\includegraphics[width=0.75\hsize]{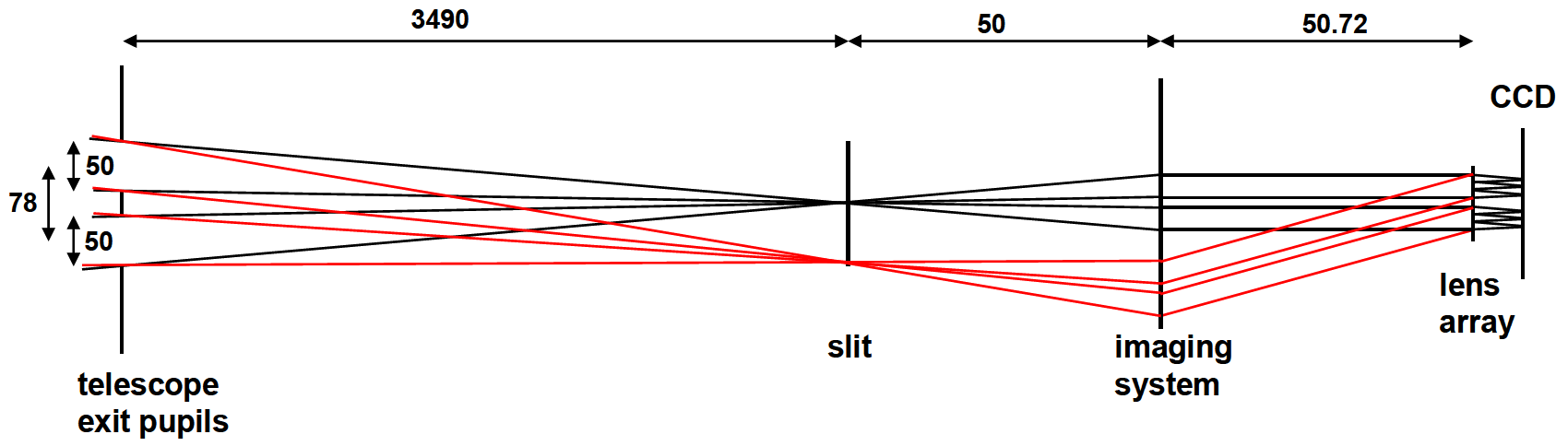}
\includegraphics[width=0.75\hsize]{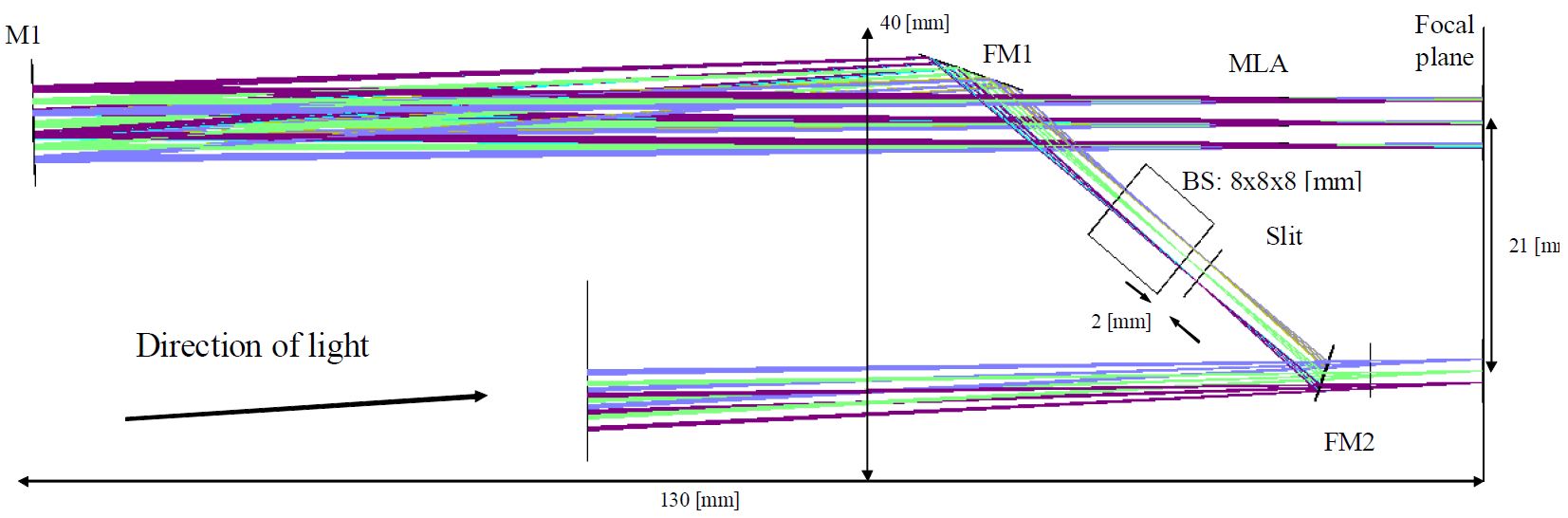}
\includegraphics[width=0.75\hsize]{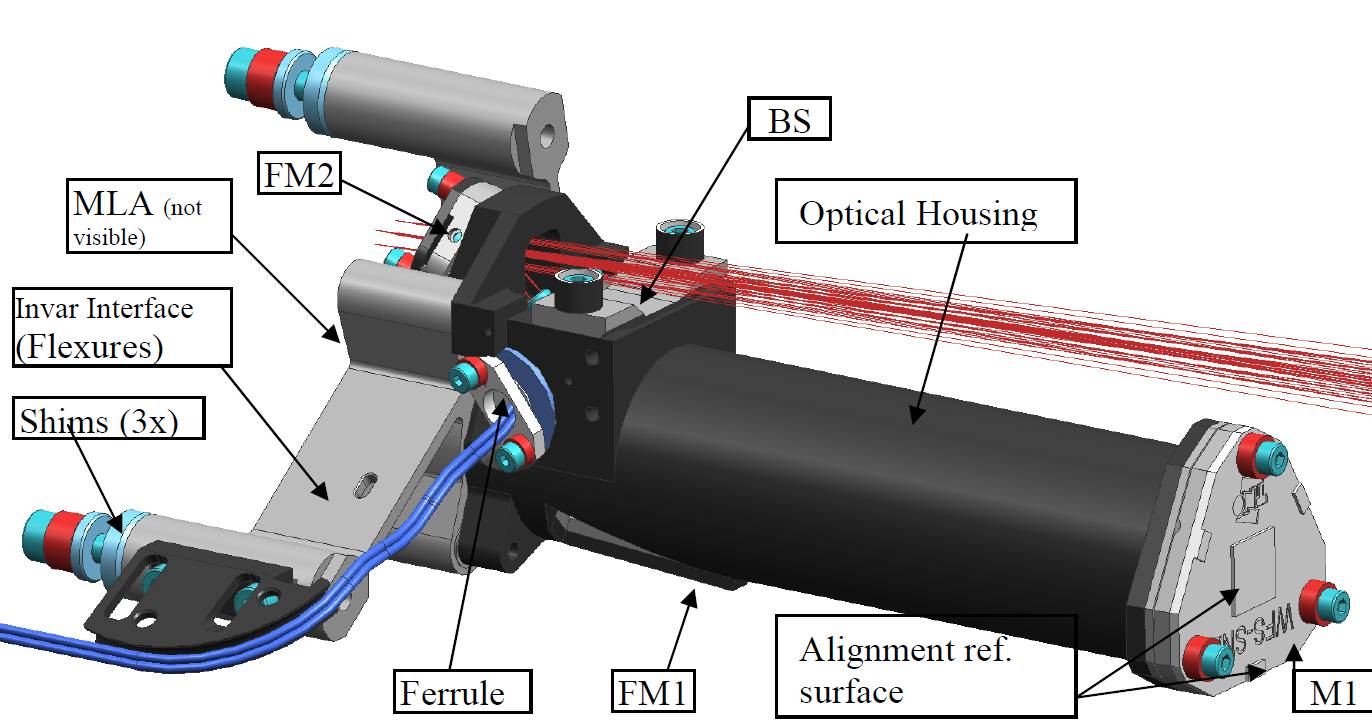}
\end{center}
\caption{Gaia wavefront sensor. Top: schematic layout of the WFS. From left to right, the gaia exit pupils and WFS entrance slit, collimator, microlens array and focal plane can be seen. Middle: optical design layout. The second fold mirror (FM2) deflects the incoming light to a beamsplitter (BS), where the reference signal from the optical fibres can be injected. The first fold mirror (FM1) sends the light to the collimator M1, which produces an image of the output pupil at the location of the microlens array (MLA). Finally, the lenslet images are projected onto the same focal plane used for the Gaia astrometric field. Bottom: mechanical overview of the WFS. Courtesy TNO\cite{2009SPIE.7439E..29V}.}
\label{fig:wfsDesign}
\end{figure}

Two Shack-Hartmann wavefront sensors , built by TNO, are located on the Gaia focal plane to provide the information required to drive the M2MM. The structure of the WFS is made of invar, while the optical surfaces are made from fused silica. Both materials provide a good thermal match to the SiC CCD support structure. The optical and mechanical design of the WFS are displayed in Fig.~\ref{fig:wfsDesign}. The former is based on an input slit (12''$\times$30''), an spherical collimator, a microlens array (387 $\mu$m pitch, 378 $\mu$m diameter), a beamsplitter cube and two fold mirrors. Each Gaia output pupil is sampled with an array of 3$\times$11 fully illuminated microlenses. Additional information on the WFS is available in the literature\cite{2009SPIE.7439E..29V}.

The analysis of the wavefront sensor is divided into three steps: centroid determination, wavefront reconstruction and M2MM actuation determination. Due to the small number of microlenses, and the low flux collected for each passing star, maximum likelihood algorithms have been developed to recover all the astrometric information contained in the WFS images. The wavefront is then approximated as a low term series of 2D Legendre polynomials, containing the first six non-trivial terms. Both processes are extensively discussed in~\cite{2012SPIE.8442E..1QM}, image centroiding reaching the maximum performance Cram\'er-Rao limit. The M2MM actuations were then derived using the Airbus telescope alignment tool, which included the Code V sensitivities for each degree of freedom and accepted individual weights for each Legendre coefficient.

\begin{figure}
\begin{center}
\includegraphics[width=0.5\hsize]{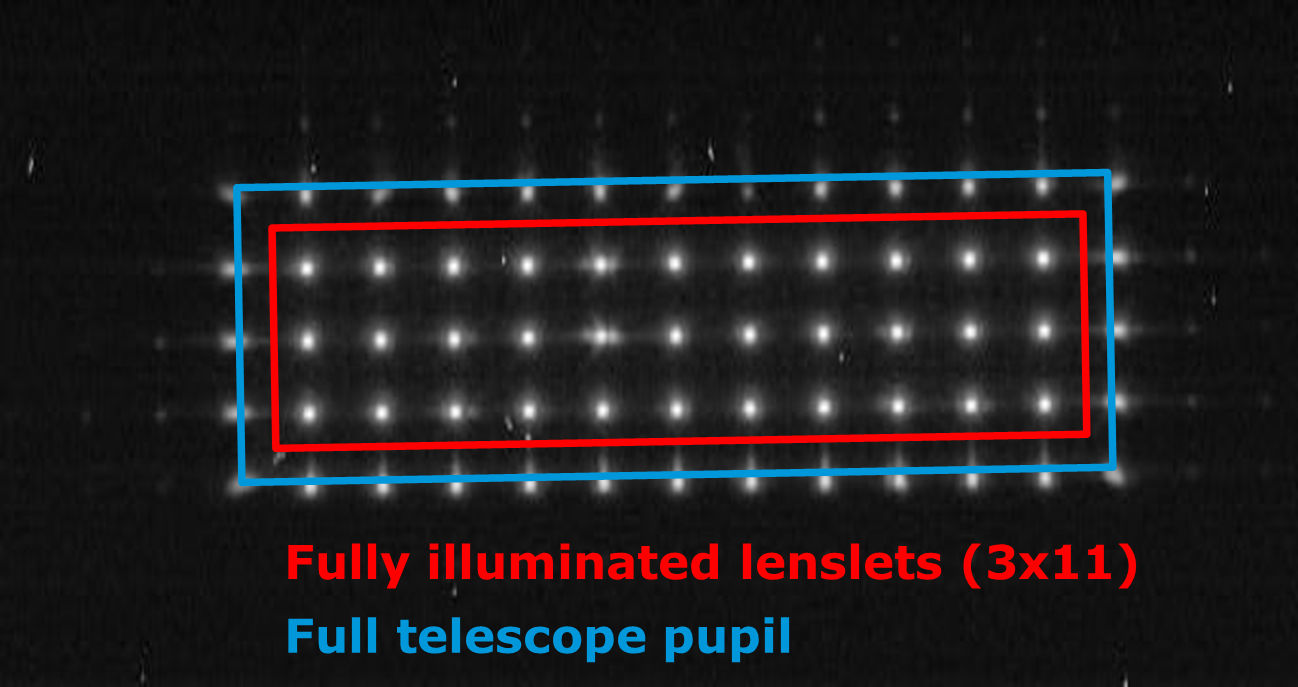}
\end{center}
\caption{WFS Stellar pattern. It includes three rows of fully illuminated microlenses surrounded by partially obscured elements.}
\label{fig:wfsStellarPattern}
\end{figure}

One significant improvement in the wavefront reconstruction compared to~\cite{2012SPIE.8442E..1QM}, is the tool developed to estimate the location and rotation of the telescope pupil with respect to the microlenses photocentre. It fits the light collected both by the fully and partially illuminated microlenses with a simple knife-edge rectangular pupil mode (see Fig.~\ref{fig:wfsStellarPattern}). The degrees of freedom are the location, rotation axis, scale factor and pupil Gaussian apodisation, the latter needed to fit on-ground test data. Significant aliasing was detected in some cases when reconstructing the wavefront with a wrong telescope pupil geometry (see Table~\ref{tab:pupilGeometryLegendre}),

\begin{table}
\caption{Influence of the telescope pupil geometry in the Legendre decomposition. Two Legendre decompositions have been carried out for some on-ground PLM TB/TV WFS1 telescope 2 patterns. For each pair, the top row assumes the pupil is aligned with the microlens array, while the bottom uses the results of a fit of the light collected by all microlenses (fully and partially illuminated). Significant differences are apparent in the L4 and L5 terms, up to 40 nm RMS.}
\label{tab:pupilGeometryLegendre}
\begin{center}
\begin{tabular}{lrrrrrr}
\hline
\hline
Test				&		L4	&		L5	&		L6	&		L7	&		L8	&		L9	\\
\hline
WFS1\_IT00	&	-61.2	&	59.1	&		0.7	&	90.1	&	63.9	&	40.8	\\
						&	-19.4	&	93.1	&		6.4	&	98.2	&	69.6	&	45.4	\\
WFS1\_IT01	&	72.8	&	54.4	&	41.3	&	77.0	&	50.4	&	39.6	\\
						&	115.3	&	82.1	&	49.2	&	82.6	&	53.8	&	43.4	\\
WFS1\_IT02	&	53.4	&	41.9	&	64.0	&	79.0	&	32.7	&	39.9	\\
						&	93.3	&	63.2	&	72.9	&	84.5	&	34.3	&	43.3	\\
\hline
\end{tabular}
\end{center}
\end{table}

\subsection{Best focus overview and science data metrics
\label{sect:bestFocusMetrics}}

\begin{figure}
\begin{center}
\includegraphics[width=0.87\hsize]{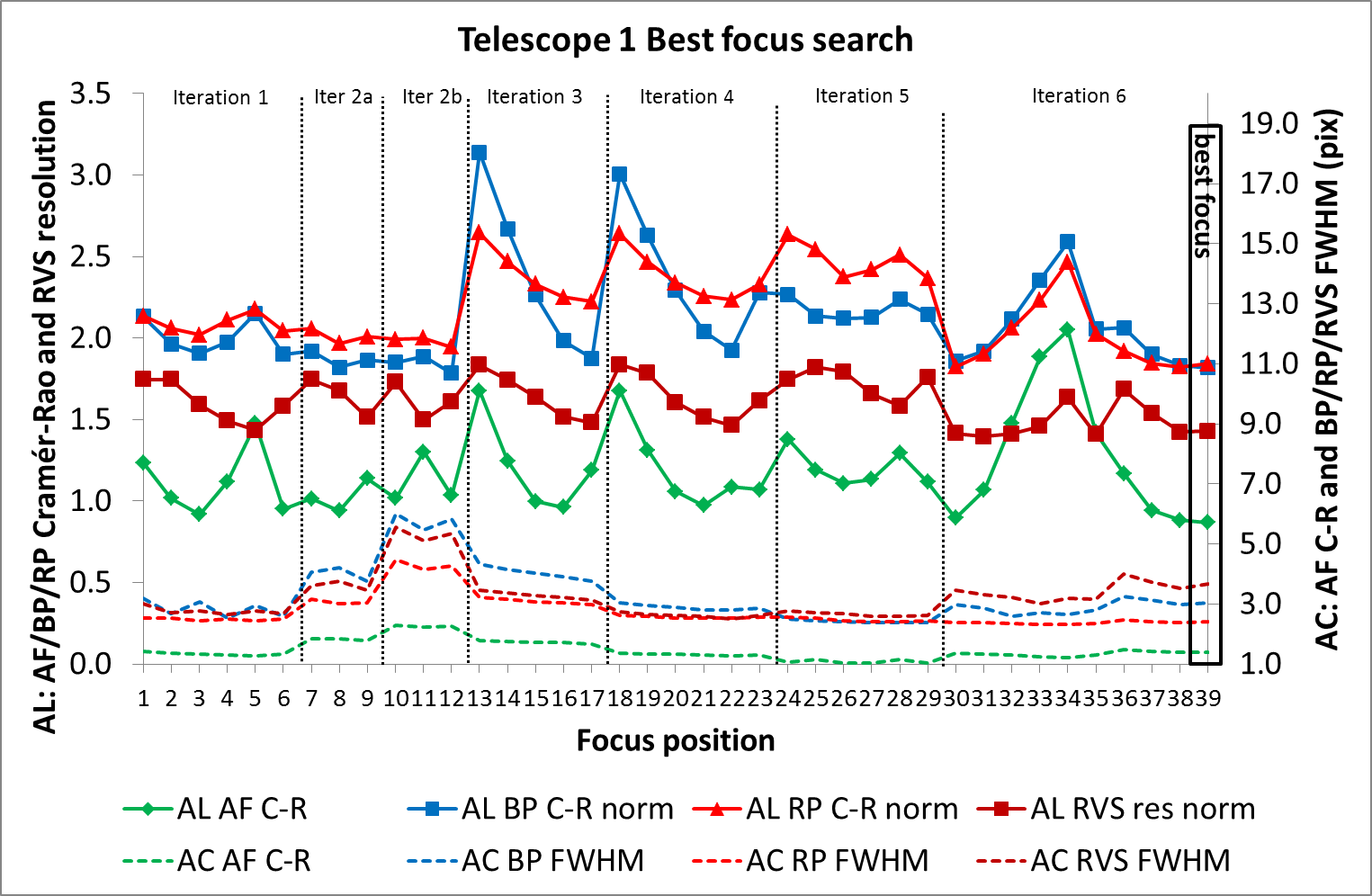}
\includegraphics[width=0.87\hsize]{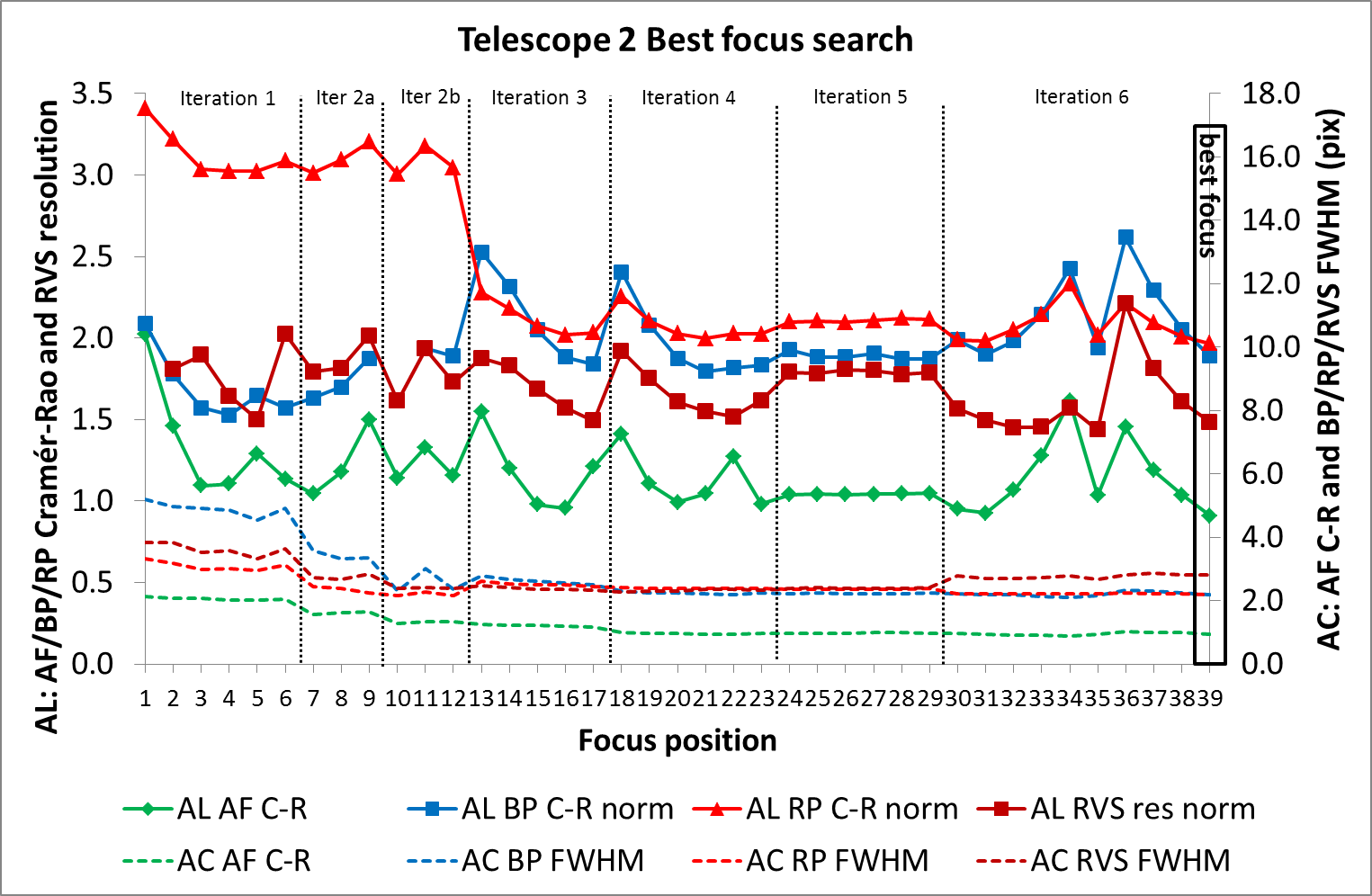}
\end{center}
\caption{Summary of the image quality assessment across the Gaia focal plane. The image quality measures for all four instruments are shown for telescope 1 (top) and 2 (bottom). The left vertical axis is the scale for the AF/BP/RP Cram\'er-Rao measures and the RVS resolving power, while the right vertical axis is the scale for the AC AF Cram\'er-Rao measure and the BP/RP/RVS AC profile widths. Purely for visualization purposes, the various BP/RP/RVS image quality measures have been normalized to bring them onto the same scale as the AF estimates. The X-axis in both panels simply indicates the 39 focus positions explored (in order), grouped by iteration (dotted vertical lines).
\label{fig:bestFocusOverview}}
\end{figure}

The metrics used to quantitatively compare the different focus positions analysed for the astrometric, photometric and spectroscopic focal planes are presented in the following sections. Fig.~\ref{fig:bestFocusOverview} presents a summary of the different focus settings. They are grouped in iterations. For each one, payload expert scientists for each focal plane provided an assessment of the optical quality, an a consensus best focus position was selected. After six iterations, convergence was found. The resultant position provides a good overall optical quality for all focal planes in the AL direction. Note that for telescope 1, the best AC image quality was obtained during iteration 5. The penalty in AL quality was too high to be accepted, though.

\subsubsection{First Look AF image quality assessment}

The image quality in AF is estimated using automatic analyses provided by the First Look system. In these diagnostics the AL centroiding performance is estimated using the Cram\'er-Rao lower bound applied to real sampled images. According to
\cite{1978moas.coll..197L,LL:2004BASNOCODE,2010ISSIR...9..279L}, the Cram\'er-Rao lower bound for single CCD AL astrometric
precision $\sigma_\eta$ is given by:

\begin{equation}
  \sigma_\eta = \frac{1}{\sqrt{ \displaystyle\sum_{k=0}^{n-1} \frac{ (S'_k)^2 }{r^2 + b + S_k} }}
\end{equation}

where $S_k$, the LSF, is the number of electrons collected from the star, binned AC, for AL pixel coordinate $k$, where $k \in [0, n-1]$. $S'_k$ is the derivative of $S_k$ with respect to the pixel coordinate, $r$ the read-out noise (in electrons) and $b$ the homogeneous sky background (in electrons). The units of $\sigma_\eta$ are pixels. They can be converted to angles (e.g. $\mu$as) multiplying by the pixel size and dividing by the telescope focal length. The following quantity is obtained for all class 0 and 1 stars:

\begin{equation}
  {\rm Cramer-Rao}_{\rm normalised}
   = \sigma_\eta \sqrt {\displaystyle\sum_{k=0}^{n-1} S_k}
   = \sigma_\eta \sqrt {N_{e^-}}
\end{equation}

The Cram\'er-Rao normalised metric is a relative measurement that does not depend on the stellar magnitude for bright objects
(stellar Poisson noise much greater than background Poisson or CCD read-out noises). The dependence on sub-pixel stellar location and colour is also small.

\subsubsection{IDT XP image quality parameter}

The IDT pipeline calculates an `image quality parameter' for the BP and RP images for all transits for which the $G$ magnitude estimated by the VPU is 16 or less. The quality parameter $\sigma_\mathrm{XP}$ is similar to the $\sigma_\eta$ parameter AF Cram\'er-Rao parameter described in the previous section:
\begin{equation}
  \sigma_\mathrm{XP} = \left(\sum_{k=l-3}^{l+3} \frac{(S^\prime_k)^2}
  {\sqrt{\sigma_\mathrm{ron}^2+\sigma_{\mathrm{bias},k}^2+\sigma_{\mathrm{back},k}^2+S_k}}
  \right)^{-1/2} \times \sqrt{S}
  \label{eq:xpimquality}
\end{equation}
The quantities in this equation are:\\
\begin{tabular}{ll}
  $S$ & total counts in BP or RP samples \\
  $S_k$ & counts in sample $k$ \\
  $S^\prime_k$ & (numerical) derivative of counts at sample $k$ \\
  $\sigma_\mathrm{ron}$ & read-out noise \\
  $\sigma_{\mathrm{bias},k}$ & error due to bias non-uniformity removal \\
  $\sigma_{\mathrm{back},k}$ & error due to background removal \\
  $l$ & index of sample containing the AL leading edge of the XP image \\
\end{tabular}

Lower values of $\sigma_\mathrm{XP}$ imply better image quality. The image quality parameter is calculated always for 1D XP images (with 2D windows summed in the AC direction) and reflects the steepness of the leading edge of 1D XP spectra. The multiplication by $\sqrt{S}$ serves to make image quality parameters for stars of different magnitudes directly comparable.

\subsubsection{RVS spectral resolution element}

The resolution element in RVS spectra represents the image quality in the AL direction. The resolution element is defined as the FWHM of an unresolved line. In the RVS wavelength range there are several Fe I lines which in most of the stars can be considered as unresolved. 

\begin{figure}
  \begin{center}
    \includegraphics[height=5cm]{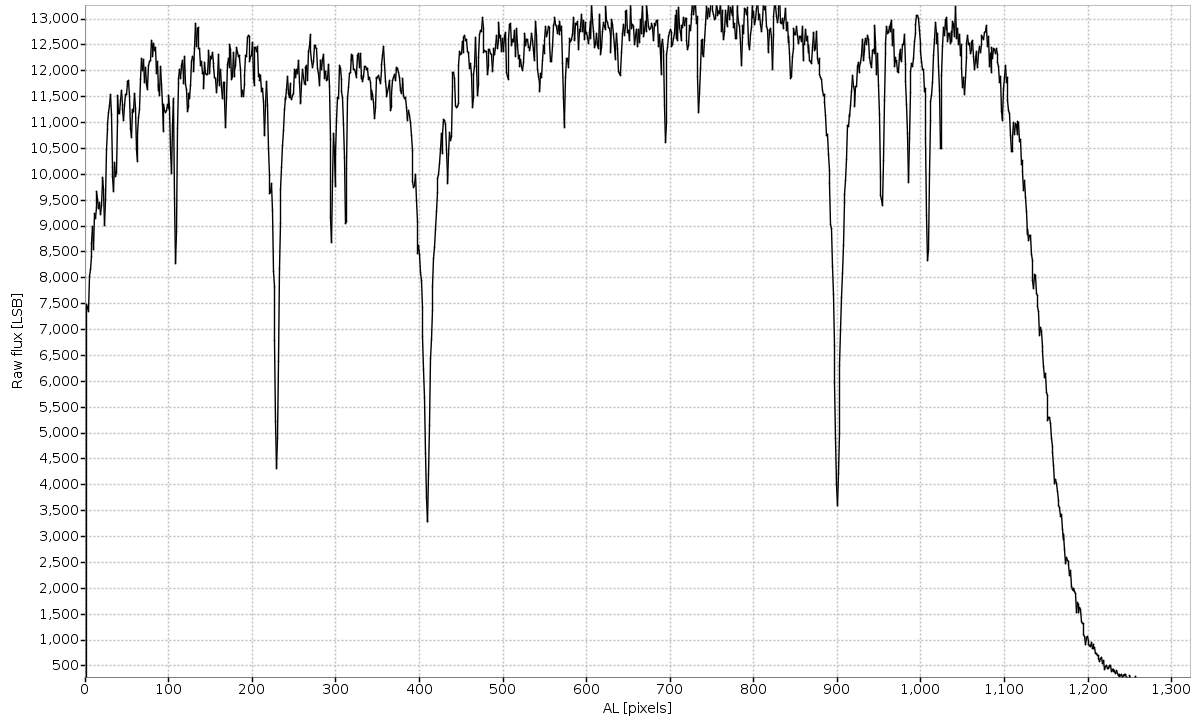}
    \includegraphics[height=5cm]{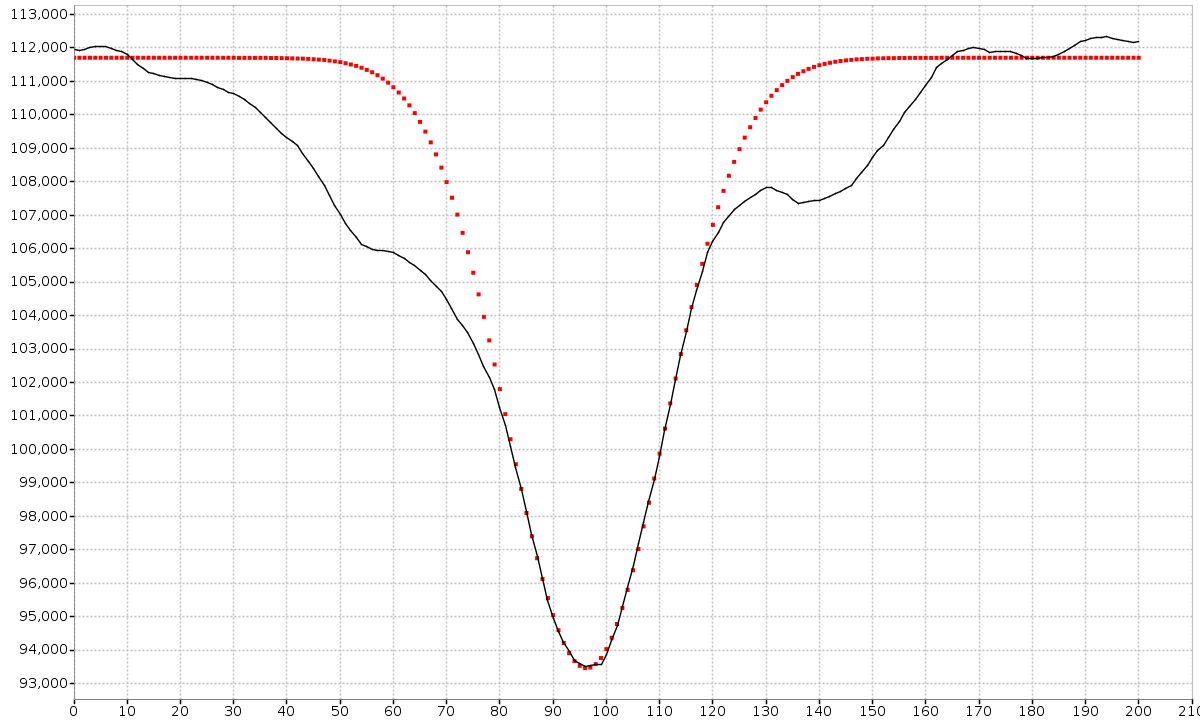}
  \end{center}
  \caption{Left panel: example of an RVS spectrum of a bright star. Right panel: the
    cross-correlation function (black) derived from the example spectrum and the fitted Gaussian
    (red dots).\label{fig:rvs_spec_corr}}
\end{figure}

The FWHM of Fe I lines has been estimated by computing the cross-correlation function (CCF) of the spectrum with a binary mask, because it can be executed automatically on a large enough number of stars. Figure \ref{fig:rvs_spec_corr} shows an example of a good measurement. This method consists in the following steps:

\begin{enumerate}
  \item subtract the bias from the spectra and collapse class 0 spectra into 1 dimension,
  \item identify the 3 Ca II lines to get an estimate of the position of Fe I lines, 
  \item produce an oversampled spectrum, 
  \item compute a binary mask corresponding to the position of 9 Fe I lines (their rest-frame
    wavelengths in air are: $851.407$, $851.511$, $852.667$, $858.226$, $861.180$, $861.628$,
    $862.160$, $867.475$, $868.863$ nm),
  \item cross-correlate the over sampled spectrum with the binary mask, 
  \item fit the cross-correlation function with a Gaussian profile
  \item obtain the resolution element as the FWHM of the fitted Gaussian profile, divided by the
    oversampling factor.
\end{enumerate}

\section{Conclusions
\label{sect:conclusions}}

Two complex problems related to the Gaia mission have been presented, related to the need of measuring the basic angle variations and refocusing the telescopes in orbit. Metrology systems have been included to tackle them: the BAM and WFSs. The latter is complemented with a detailed analysis of the scientific data to define and obtain the best focus performance.

The main conclusion regarding the BAM is that it works. This is remarkable, because this system is the highest precision interferometer ever flown to space. The BAM data display three major features: a) a periodic signal, which can be characterised at the $\mu$as level, b) discontinuities, some of which have been identified as real sudden changes of the differential telescope line of sight and c) long term trends in the scale of days and weeks, which have been proven to be artificial, but irrelevant, because the self-calibrating data analysis will filter those false trends. Finally, the BAM has displayed a very high sensitivy to any significant disturbance of normal spacecraft operations, which makes it a very convenient alert tool.

Gaia has also been successfully focused. Both wavefront sensors have worked up to the expectations. However, they alone are not enough to define the scientific best focus. Ad-hoc data analysis has been carried out in the four Gaia focal planes (astrometric, blue and red photometric and spectroscopic). The final focus settings have been chosen as a compromise to provide good performance for all parties. The image quality achieved is, in any case, remarkable, and the refocusing activity a success.

\section{Acknowledgements}

The authors wish to thank Airbus Defence \& Space and TNO for their support and access to internal documents on the wavefront sensor and Gaia optical design. Some concepts and ideas presented here come from those sources.

Material used in this work has been provided by the Coordination Units 3, 5 and 6 (CU3, CU5, CU6) of the Gaia Data Processing and Analysis Consortium (DPAC). They are gratefully acknowledged for their contribution.

\bibliography{2014_06_spie_metrology,gaia_livelink_valid,gaia_livelink_obsolete,gaia_drafts,gaia_refs,gaia_books,gaia_refs_ads}   %>>>> bibliography data in report.bib
\bibliographystyle{spiebib}   %>>>> makes bibtex use spiebib.bst

\end{document}